\documentclass{optica-article}

\journal{opticajournal} % for journals or Optica Open

\articletype{Research Article}

\begin{document}

\title{Simultaneous Charge Carrier Density Mapping of SiC Epilayers and Substrates with Terahertz Time-Domain Spectroscopy}

\author{Joshua Hennig\authormark{1,2,*}, Jens Klier\authormark{1}, Stefan Duran\authormark{1}, Kuei-Shen Hsu\authormark{3}, Jan Beyer\authormark{3}, Christian Röder\authormark{4}, Franziska C. Beyer\authormark{4}, Nadine Schüler \authormark{5}, Nico Vieweg\authormark{6}, Katja Dutzi\authormark{6}, Georg von Freymann\authormark{1,2}, and Daniel Molter\authormark{1}}

\address{\authormark{1}Fraunhofer Institute for Industrial Mathematics ITWM, Department Materials Characterization and Testing, 67663\,Kaiserslautern, Germany\\
\authormark{2}Department of Physics and Research Center OPTIMAS, RPTU University Kaiserslautern-Landau, 67663\,Kaiserslautern, Germany\\
\authormark{3}Institute of Applied Physics, Technische Universität Bergakademie Freiberg, Leipziger Str. 23, 09599\,Freiberg, Germany\\
\authormark{4}Fraunhofer Institute for Integrated Systems and Device Technology IISB, Department Energy Materials and Test Devices, Schottkystraße 10, 91058\,Erlangen, Germany\\
\authormark{5}Freiberg Instruments GmbH, Delfter Str. 6, 09599\,Freiberg, Germany\\
\authormark{6}TOPTICA Photonics AG, Lochhamer Schlag 19, 82166\,Gräfelfing, Germany\\}

\email{\authormark{*}joshua.hennig@itwm.fraunhofer.de}

\begin{abstract*}  

With the growing demand for efficient power electronics, SiC-based devices are progressively becoming more relevant. In contrast to established methods such as the mercury capacitance-voltage technique, terahertz spectroscopy promises a contactless characterization. In this work, we simultaneously determine the charge carrier density of SiC epilayers and their substrates in a single measurement over a wide range of about 8$\times$10$^{15}$ cm$^{-3}$ to 4$\times$10$^{18}$ cm$^{-3}$ using time-domain spectroscopy in a reflection geometry. Furthermore, inhomogeneities in the samples are detected by mapping the determined charge carrier densities over the whole wafer. Additional theoretical calculations confirm these results and provide thickness-dependent information on the doping range of 4H-SiC, in which terahertz time-domain spectroscopy is capable of determining the charge carrier density.

% With the growing demand for efficient power electronics due to the expansion of renewable energies and the electrification of all branches of mobility, silicon carbide-based electronic devices are becoming more and more relevant. The established method for measuring the charge carrier density of SiC epilayers is the contact-based mercury capacitance-voltage technique. Terahertz time-domain spectroscopy has been proven to be suitable for characterizing optical and electrical properties of semiconductors in a non-contact manner. Therefore, we calculate simulations based on the Drude model to determine the doping range of 4H-SiC, TDS can measure the charge carrier density for. Measurements of multilayer SiC samples prove these simulations and show the possibility of measuring both the charge carrier density of the epilayer and the substrate in a single measurement, over a wide range of about 8$\times$10$^{15}$ to 4$\times$10$^{18}$ cm$^{-3}$. Furthermore, the fast measurement rate of an ECOPS system is used to map the charge carrier density with a resolution of 1 mm across 6" wafers allowing to detect inhomogeneities in the samples. With these results, TDS appears to be a promising and capable technique in quality and production control of 4H-SiC to help satisfy the demand for SiC-based power devices in the future.

\end{abstract*}

%%%%%%%%%%%%%%%%%%%%%%%%%%  body  %%%%%%%%%%%%%%%%%%%%%%%%%%
\section{Introduction}
Silicon carbide (SiC) is a modern semiconductor material that has become established for certain power devices such as diodes or transistors for specific applications over the last three decades \cite{she2017review, han2020review, langpoklakpam2022review}. Perspectively, it is expected to become even more applied in further fields, e.g., in electric vehicles, electric train traction, wind turbines, or photovoltaic inverters \cite{buffolo2024review, thoma2018design}. Therefore, it plays an important role in power electronic devices and energy converters, which are necessary for the energy transition. The advantages of SiC-based devices over silicon-based devices are their increased thermal stability due to a higher melting point and thermal conductivity, as well as an increased robustness and compatibility to higher voltages due to a higher breakdown electric field and a wider bandgap \cite{she2017review, alves2017sic}. With the further growing need for devices in demanding environments, that can benefit from these advantages, technologies to characterize relevant properties such as the charge carrier density of doped SiC epilayers are necessary.

The established technique to measure the charge carrier density of SiC epilayers is the mercury capacitance-voltage (mCV) technique \cite{mizsei2014characterization, schroder2015semiconductor, sanna2021assessment}.
% It makes use of the fact that a contact between a drop of Hg and a semiconductor can form a Schottky barrier with a reversed-biased space-charge region. The width of this space-charge region depends on the work function of both materials and on the applied voltage. By applying a small AC voltage, the capacitance can be determined. In the characteristic mCV formula, the charge carrier density depends antiproportionally on the relative permittivity of the semiconductor and on the square of the contacted area \cite{schroder2015semiconductor, sanna2021assessment}. 
Unlike other contact-based measurement techniques used to measure the resistivity, such as the four-point probe method \cite{smits1958measurement, naftaly2021sheet}, mCV is used to directly measure the charge carrier density \cite{schroder2015semiconductor}. However, mCV is also a contact-based method, with certain disadvantages coming along with it. These include that contact with another material can potentially affect or damage the semiconductor material under test and the measurement technique is comparably slow due to the time necessary to prepare the contact. Additionally, there is the problem to remove the mercury from the surface residue-free, thus wafers can be further processed without contamination issues.
% Furthermore, working with mercury is not desirable due to its toxicity requiring special carefulness and additionally the wafer measured should be cleaned  after the measurement. 
These drawbacks, together with a strong dependence on the contacted area as well as potentially necessary surface preparation \cite{sanna2021assessment}, illustrate the need of a non-contact measurement technique for the charge carrier density from both an industrial as well as a scientific point of view.

One alternative method to access the doping concentration optically is Raman spectroscopy. Based on the frequency shift of the LO phonon plasmon coupled (LOPC) mode due to the change in charge carrier density in doping ranges above 2$\times10^{16}$ cm$^{-3}$ \cite{nakashima2008determination}. Another non-contact method is the fairly new corona-charge non-contact capacitance voltage technique (CnCV). It requires a corona discharge to precisely deposit charge increments on the surface of the sample, which is placed on a movable conducting chuck, held in place by vacuum. In a next step, the sample is moved to a Kelvin probe, where the CV measurement is performed in a scanning mode. From this measurement, the charge carrier density can be retrieved. Afterwards, the additional corona charges are removed by UV illumination \cite{wilson2022characterization}. However, both of these methods are comparably complex and therefore not yet established as standard techniques.

Terahertz time-domain spectroscopy (TDS) enables a measurement of the charge carrier density that is completely non-contact, allows measurements under ambient conditions, and is potentially faster than mCV. Since the first realizations of photoconductive antennas and the appearances of TDS \cite{auston1984picosecond ,cheung1986novel, van1989terahertz}, it has found applications, e.g., in non-destructive material testing and characterization such as layer thickness measurements\cite{jen2014sample, krimi2016highly, klier2021thickness}, defect detection \cite{ellrich2020terahertz, ospald2014aeronautics}, spectroscopy of dangerous or illegal substances \cite{davies2008terahertz, molter2021mail}, measurements of dielectrics \cite{wietzke2010terahertz, ma2019dielectric}, and historical art analysis \cite{fukunaga2010terahertz}. % , picollo2015obtaining
%, and many more \cite{naftaly2007terahertz, naftaly2019industrial, ren2019state, afsah2019comprehensive, bawuah2021advances, jepsen2011terahertz, neu2018tutorial}. 
Remarkably, the potential of TDS to characterize electrical properties such as the resistivity of semiconductor materials had been demonstrated already in early pioneering works \cite{van1990optical, katzenellenbogen1992electrical}. This branch of terahertz research has continued to advance since then, leading to the analysis of highly doped thin films \cite{ulatowski2020terahertz} and two-dimensional materials \cite{mitra2025terahertz} such as graphene \cite{boggild2017mapping}. Although the classical approach of TDS to measure in transmission geometry is widely used, measuring in reflection geometry allows one to retrieve information from optically dense samples that are not transparent to terahertz radiation \cite{jeon1998characterization, nashima2001measurement, hennig2024wide}.
This is the case for doped 4H-SiC epilayers which are typically grown on highly doped 4H-SiC substrates by chemical vapor deposition \cite{kallinger2013step, matsunami1997step, syvajarvi1999growth}. While for example for GaN, the characterization of epilayers with TDS had been demonstrated before \cite{guo2009terahertz}, this has -- to the best of our knowledge -- not been achieved for SiC, yet.

% This branch of terahertz research has been advancing further leading to measurements of electrical properties of semiconductor nanowires \cite{joyce2016review}, the analysis of highly doped thin films \cite{ulatowski2020terahertz} and two-dimensional materials \cite{mitra2025terahertz} such as graphene \cite{boggild2017mapping}. 

\section{4H-SiC model}

Silicon carbide has a large number of more than 250 polytypes \cite{cheung2006silicon}. However, only a small number of them are commonly used, including 4H-SiC, which has multiple industrial applications\cite{she2017review}. This polytype has a hexagonal crystal structure, as the name suggests. 

\begin{figure}[htbp!]
\centering
\includegraphics[width=0.12\textwidth]{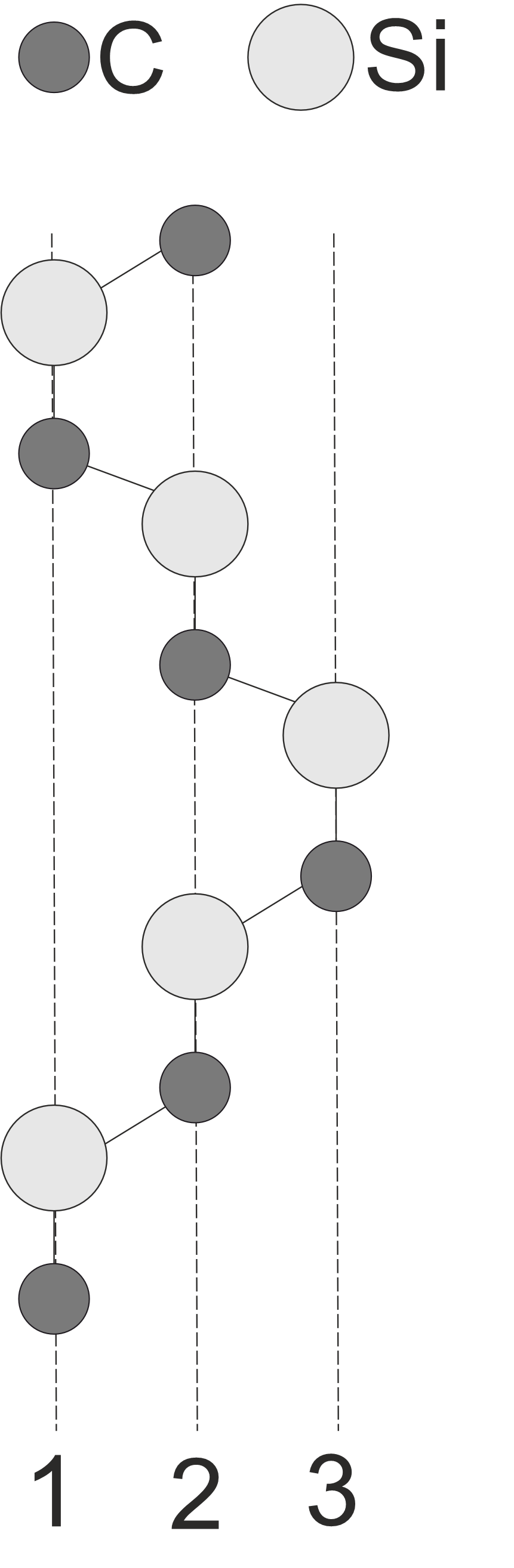}
\put(-70,140){(a)}
\hspace{3cm}
\includegraphics[width=0.2\textwidth]{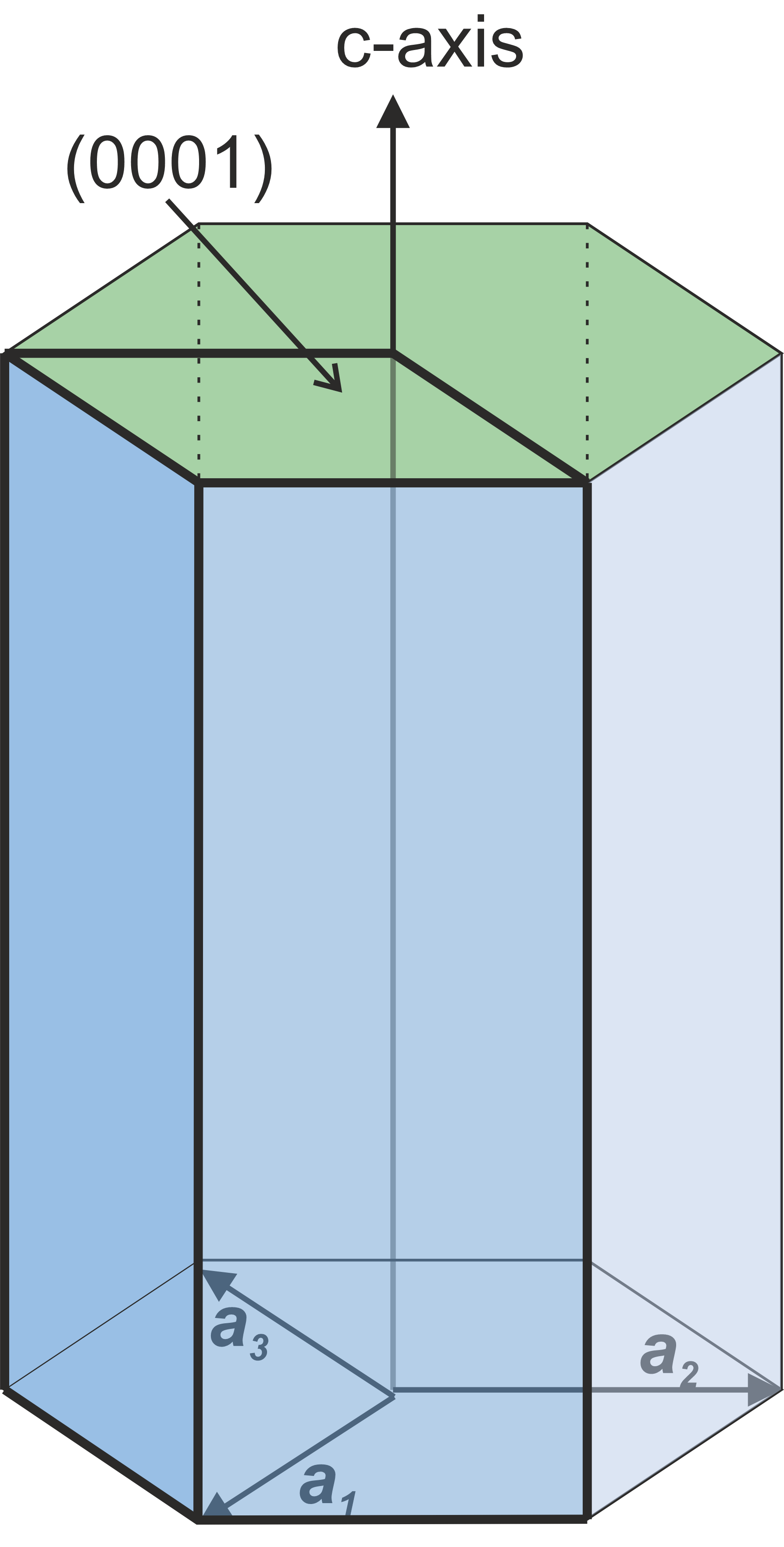}
\put(-100,140){(b)}
\caption{(a) Crystal structure of 4H-SiC showing the stacking sequence of C and Si atoms, forming a Si-C double layer. 1, 2, and 3 denote the occupation sites. (b) Unit cell (thick lines) in the hexagonal structure of 4H-SiC with the three translation vectors and the c-axis. Sketches inspired by \cite{matsunami2004technological, wang2020density, kimoto2022high}} 
\label{fig:SiC_unit_cell}
\end{figure}

The crystal structure of 4H-SiC is sketched in Fig. \ref{fig:SiC_unit_cell} (a). The numbers 1, 2, and 3 denote the different occupation sites for a Si-C double layer. After four Si atoms and four C atoms, the structure is repeated. Fig. \ref{fig:SiC_unit_cell} (b) shows a unit cell of the hexagonal crystal structure, highlighted by the thick lines. The translation vectors $\boldsymbol{a}_{1}$, $\boldsymbol{a}_{2}$, and $\boldsymbol{a}_{3}$ span a coordinate system. The (0001) plane is marked in green and perpendicular to the c-axis of the hexagonal material.

The optical and electrical properties of a semiconductor, e.g., Si can be described by the Drude model \cite{nashima2001measurement}. This also holds for SiC \cite{agulto2025wafer}. In the Drude model, the complex relative permittivity $\tilde{\epsilon}$ is given by

\begin{equation}
    \tilde{\epsilon} = \epsilon_{\text{S}}-\frac{\omega_{\text{p}}^{2}}{\omega(\omega+\text{i}\Gamma)}
    \label{eq:dielectric_function_Drude}
\end{equation}

with $\epsilon_{\text{S}}$ being the static relative permittivity, $\omega_{\text{p}}$ the plasma frequency , $\omega$ the angular frequency, $i$ the imaginary unit and $\Gamma = 1/\tau$ the plasmon damping frequency. $\Gamma$ is the inverse of the mean free time $\tau = \frac{m_{\text{eff}}m_{\text{e}}\mu}{e}$ and the square of the plasma frequency can be expressed by $\omega_{\text{p}}^{2} = Ne^{2}/(\epsilon_{0}m_{\text{eff}}m_{\text{e}})$. Here, $N$ is the carrier density, $e$ is the elementary charge, $\epsilon_{0}$ the vacuum permittivity, $m_{\text{e}}$ the electron mass, $m_{\text{eff}}$ the effective mass for conductivity and $\mu$ the charge carrier mobility. For the mobility, an empirical model based on the description proposed for Si by Caughey and Thomas \cite{caughey1967carrier} can also be used for SiC

\begin{equation}
    \mu(N) = \frac{\mu_{\text{max}} - \mu_{\text{min}}}{ 1 + \big (\frac{N}{N_{\text{ref}}} \big )^{\beta}} + \mu_{\text{min}} .
    \label{eq:mobility_empirical_formula}
\end{equation}

In Eq. (\ref{eq:mobility_empirical_formula}) the specific properties $\mu_{\text{max}}$, $\mu_{\text{min}}$, $N_{\text{ref}}$ and $\beta$ are assumed as constant values. In addition, the temperature dependence could be taken into account by scaling the parameters $\Theta$ with the temperature according to $\Theta = \Theta_{300}(T/300)^{\xi}$ \cite{burin2024tcad}.  The exponent $\beta$ corresponds to the exponent $\alpha$ in the original work \cite{caughey1967carrier}. This renaming is done to avoid confusion with the extinction coefficient $\alpha$. This formula for $\mu$ is substituted into the formula for $\tau$ or $\Gamma$, respectively, which is inserted into Eq. (\ref{eq:dielectric_function_Drude}) together with $\omega_{\text{p}}$ leading to the following expression for $\tilde{\epsilon}$:

\begin{equation}
    \tilde{\epsilon}(N, \omega) = \epsilon_{\text{S}}-\frac{Ne^{2}}{\epsilon_{0}m_{\text{eff}}m_{\text{e}}\omega \big (\omega+\text{i}\frac{e}{m_{\text{eff}}m_{\text{e}}\mu(N)} \big )}.
    \label{eq:dielectric_function_Drude_N_dependend}
\end{equation}

Eq. (\ref{eq:dielectric_function_Drude_N_dependend}) has the benefit over Eq. (\ref{eq:dielectric_function_Drude}) that it has a direct dependence only on $N$ and $\omega$, which will be advantageous in the later evaluation. The other parameters of Eq. (\ref{eq:dielectric_function_Drude_N_dependend}) are considered to be material constants that can be taken from literature for the sample under test. In case of an incident electric field  perpendicular to the c-axis, the corresponding perpendicular effective mass with the formula $m_{\text{eff,}\perp} = \frac{2m_{M\Gamma}m_{MK}}{2m_{M\Gamma}+m_{MK}}$ is used \cite{ishikawa2023experimental}. Here, $m_{M\Gamma}$ and $m_{MK}$ are the effective masses along the corresponding axes in the Brillouin zone between the edge center $M$ and the center of the Brillouin zone $\Gamma$ or the middle of an edge $K$, respectively.

With the complex relative permittivity, the real and imaginary parts of the complex refractive index can be calculated using the two formulas in Eq. (\ref{eq:eps_to_n}). Furthermore, using the Fresnel equations, the reflectance and the transmittance at an interface can be calculated. 

\begin{equation}
    n_{r} = \frac{1}{\sqrt{2}}\sqrt{\epsilon_{r}+(\epsilon_{i}^{2}+\epsilon_{r}^{2})^\frac{1}{2}} \;\;\;, \;\;\; n_{i} = \frac{1}{\sqrt{2}}\sqrt{-\epsilon_{r}+(\epsilon_{i}^{2}+\epsilon_{r}^{2})^\frac{1}{2}}.
    \label{eq:eps_to_n}
\end{equation}

\begin{figure}[htbp!]
\centering
% \includegraphics[width=0.49\textwidth]{Images/Drude/SiC_Drude_Simulations_fp_N.png}
% \put(-190,120){(a)}
% \hspace{0.1cm}
% \includegraphics[width=0.49\textwidth]{Images/Drude/SiC_Drude_Simulations_epsilon_real_N.png}
% \put(-190,120){(b)}

% \vspace{0.01cm}
\includegraphics[width=0.49\textwidth]{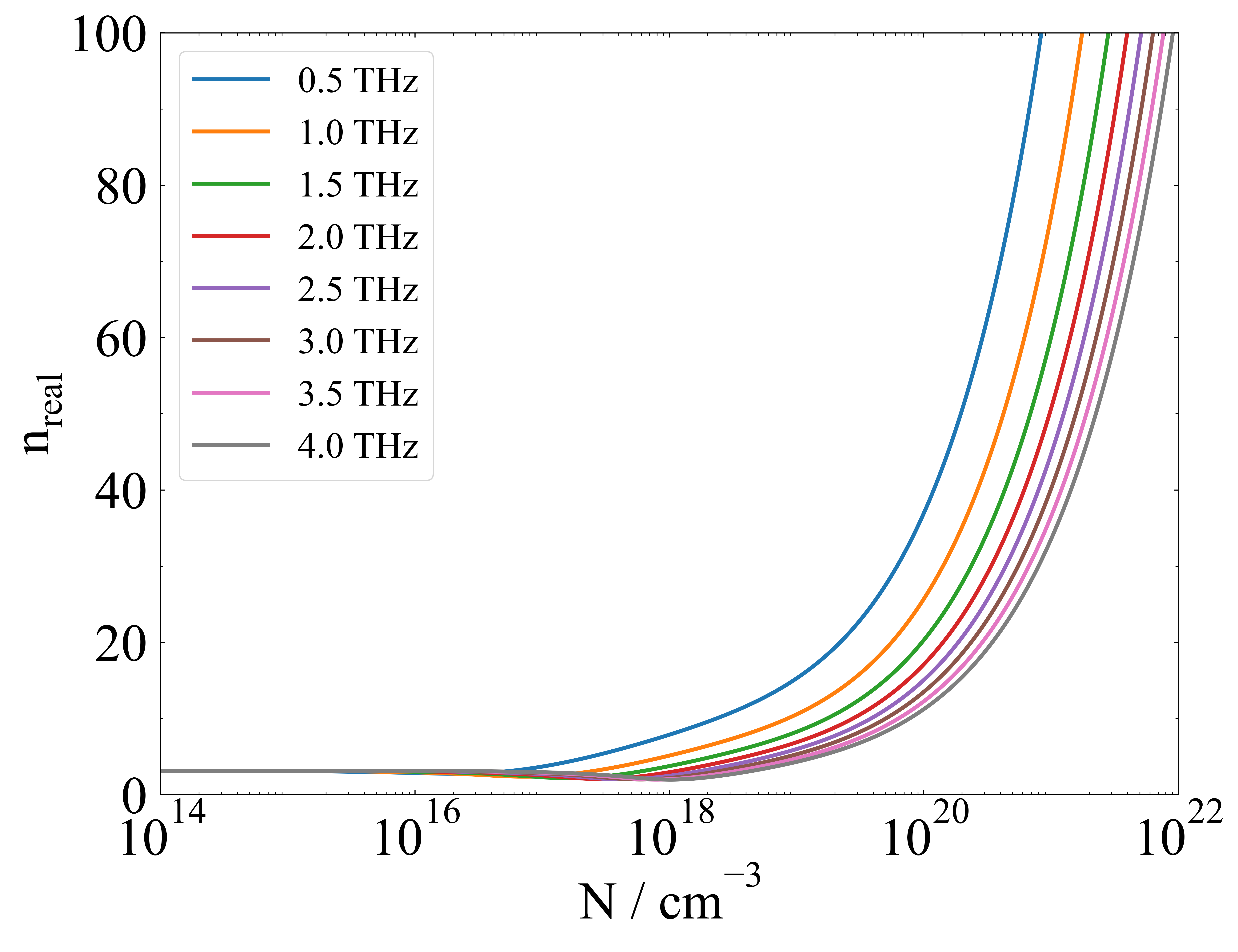}
\put(-190,120){(a)}
\hspace{0.1cm}
\includegraphics[width=0.49\textwidth]{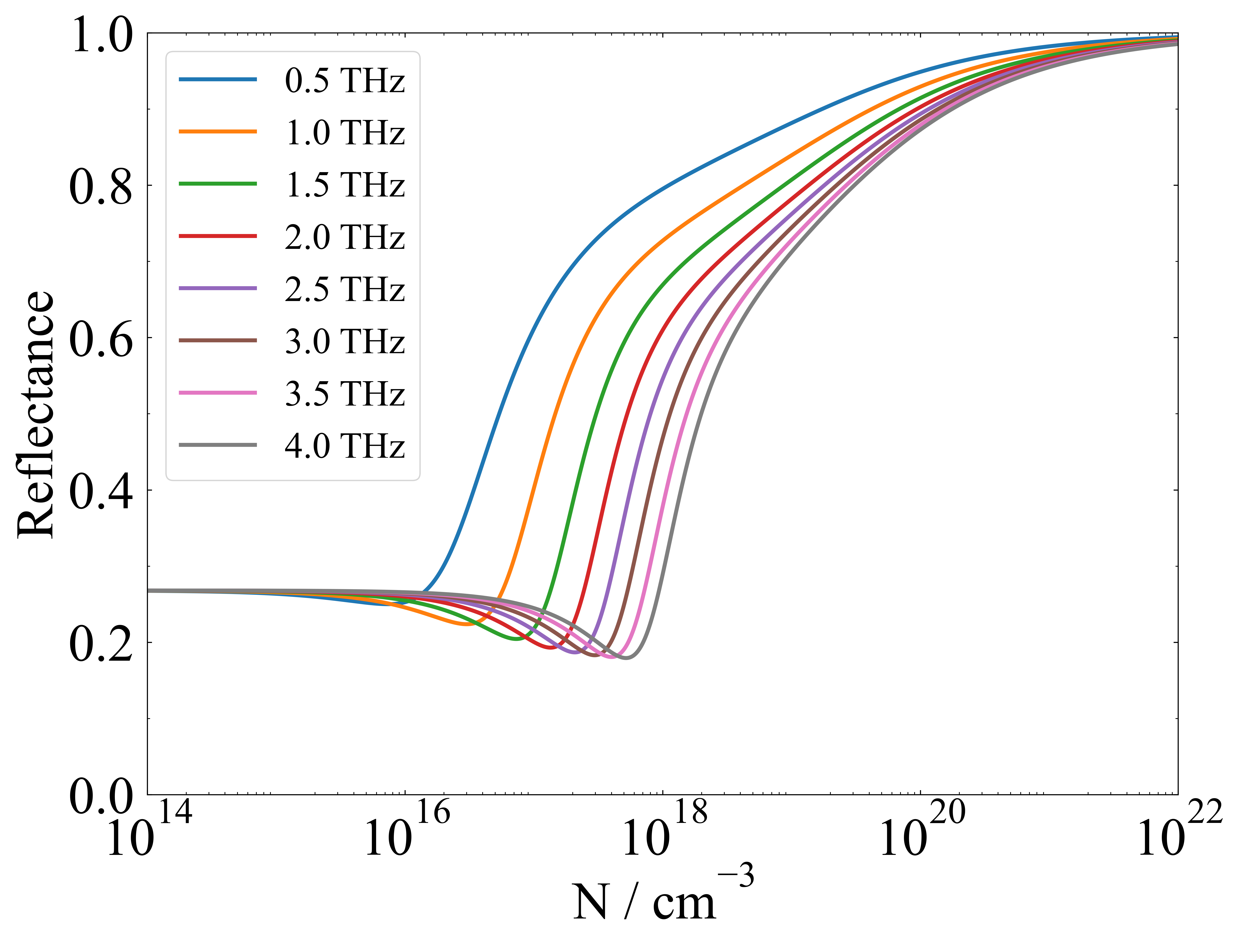}
\put(-190,120){(b)}
\caption{Theoretical calculations of (a)  the real part of the refractive index, and (b) the reflectance depending on the charge carrier density for 4H-SiC. The curves are plotted in different colors for the terahertz frequencies relevant for the TDS measurements.}
\label{fig:SiC_Drude_model}
\end{figure}

With this model, we theoretically calculate the expected material properties. Figs. \ref{fig:SiC_Drude_model} (a) and (b) show the real part of the refractive index and the reflectance in dependence of $N$, respectively. The graphs are shown for different frequencies between 0.5 THz and 4.0 THz, which is the relevant range for TDS measurements. With an increase in charge carrier density, the plasma frequency also increases. For that reason, the refractive index and finally the reflectance, which is the directly measured quantity, show most variation in the range from around 5$\times$10$^{15}$ cm$^{-3}$ to 10$^{21}$ cm$^{-3}$, according to Eq. (\ref{eq:eps_to_n}), and the Fresnel equations. Within this doping range, TDS measurements likely exhibit the greatest dynamic range and can accurately determine the charge carrier density of the material. This effect is examined in more detail in the simulations of Chapter 5. %Therefore, this doping range is where one can expect TDS measurements to allow to retrieve the charge carrier density of the material because it exhibits the most dynamic, which is crucial for the iterative determination of the carrier density as explained later in chapter 4. The farther away from this doping density, the less dynamic is to be expected in the measurement, as simulated in more detail in chapter 5.

\section{Measurement setup}

In order to achieve a fast measurement, a commercially available TDS system based on electronically controlled optical sampling (ECOPS), the TOPTICA TeraFlash smart system \cite{TOPTICA_TeraFlash_smart}, is used. Photoconductive antennas \cite{kohlhaas2022ultrabroadband} serving as emitter and detector are positioned in the measurement head, which is set up for reflection measurements. This system can reach measurement rates of up to 1600 Hz \cite{yahyapour2019fastest}. 
% This specific setup of a TDS system is based on two separate fs-lasers. One of them is operated at a fixed repetition rate while the repetition rate of the other one is slightly modulated. This way, a phase difference between the two pulses is realized \cite{kim2010high, dietz2014all, yahyapour2019fastest}. This replaces the mechanical delay stage, which is the common way to induce a phase shift between the emitter and the detector arm of a TDS system \cite{neu2018tutorial}.

\begin{figure}[htbp!]
\centering
\centering\includegraphics[width=0.75\textwidth]{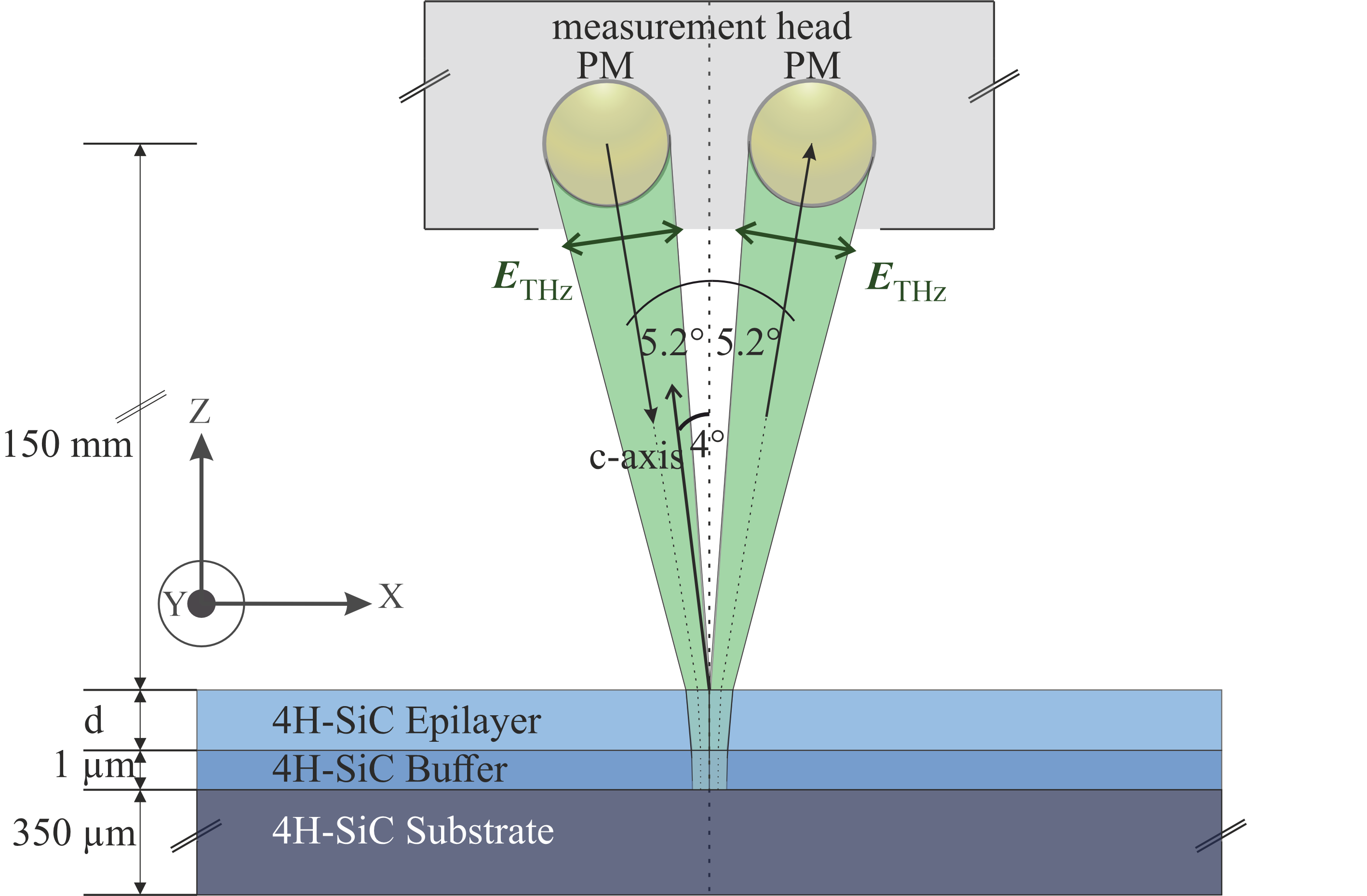}
\caption{Sketch of the setup of the measurement head above the sample with the layer sequence of the 4H-SiC samples shown in tones of blue and terahertz beam paths indicated in green. The angles under which the terahertz light is reflected from the parabolic mirrors and the angle of the c-axis of the material as well as the polarization of the light are sketched. The sketch is not to scale.} 
\label{fig:Setup}
\end{figure}

The setup of the sample under the measurement head in reflection geometry is sketched in Fig. \ref{fig:Setup}. The measurement head with the two parabolic mirrors (PM) is shown on top. One of them focuses the collimated terahertz light, which originally comes from an emitter inside the measurement head, onto the sample. The other one guides the light from the sample to the detector, each at an angle of approx. 5.2°. The terahertz radiation is indicated with a linearly polarized electric field $\boldsymbol{E}_{\text{THz}}$. The 4H-SiC sample consists of three layers, a highly nitrogen-doped substrate, a slightly less doped buffer layer, and an even weaker doped epilayer. In the case of a lightly doped epilayer, this layer is transparent for the terahertz radiation. The highly doped substrate is not transparent, yet it is still possible to measure it through the also highly doped, but only 1 \textmu m thin buffer layer. The 4°-off axis orientation of the material's c-axis, which is beneficial for a 4H-SiC growth with little defects \cite{chen2005growth}, is also shown. 

In this work, 100 measurements are averaged to acquire each data point.
% Due to the high measurement rate of the ECOPS system, averaging multiple measurements and thereby increasing the signal-to-noise ratio is possible while still acquiring measurements fast. For the data points shown in this paper, 100 measurements were averaged.
%Imaging with TDS is typically achieved by raster scanning and mapping the desired quantity to the position of the measurement \cite{boggild2017mapping, mittleman1997noncontact, bouchard2022terahertz, castro2022recent}. 
Two translation stages are used to move the measurement head continuously laterally above the wafer both in X- and Y-direction to image the whole wafer, keeping the distance to the sample constant at the focal length of the parabolic mirrors of 150 mm.  At the focal point of the PM, the spot size is about 1 mm in diameter, depending on the frequency considered. The distance between each measurement point is 1 mm. 

\section{Data evaluation}

Although the pulsed electric field measurement is recorded in the time domain, dispersion must be considered during evaluation to utilize all available phase and amplitude information in the TDS measurement. Therefore, a fast Fourier transform (FFT) is applied to switch to the frequency domain, resulting in a complex, frequency-resolved spectrum.

The reflectance in Fig. \ref{fig:SiC_Drude_model} (b) is calculated for different frequencies, yet it is just the description of the effect of a single interface between air and SiC. To describe the full effect on a terahertz pulse, that a multilayer sample consisting of an epilayer, a buffer layer, and a substrate has, a stack of layers needs to be modeled, including every layer of the material and its respective material properties. For this purpose, Rouard's method is used as a numeric approach \cite{krimi2016non, vasicek1950reflexion}. In this method, the transfer function $H$ describes the effect of the sample on the electric field in the frequency domain. To obtain a pulse in the time domain, an inverse Fourier transform $\mathscr{F}^{-1}$ needs to be applied to the result.

\begin{equation}
    E(t) = \mathscr{F}^{-1}[H(\omega)\mathscr{F}[E_0(t)]]
    \label{eq:Rouard_transfer_function_E_field}
\end{equation}

The transfer function for the case of reflection at a single layer with the layer number $l$ is given by a combination of the involved Fresnel coefficients for transmission $t$ and reflection $r$ as well as an exponential term describing the phase change and absorption losses

\begin{equation}
    H_{l}(\omega) = r_{l-1, l} + \frac{t_{l-1, l}\, r_{l, l+1}\, t_{l, l-1}\, \text{exp} \big (\frac{2\text{i}\tilde{n_{l}}\omega d_{l}}{c} \big )}{1-r_{l, l-1}\, r_{l, l +1}\, \text{exp} \big (\frac{2\text{i}\tilde{n_{l}}\omega d_{l}}{c} \big )}.
    \label{eq:Rouard_transfer_function_single}
\end{equation}

Here, the index $l > 0$ refers to the layers of different materials and $l = 0$ is the air above the sample. In the case of a multilayer system, the first layer modeled by Eq. (\ref{eq:Rouard_transfer_function_single}) is the substrate layer. Then, the next layers are modeled analogously and instead of using the reflection coefficient between the layers $r_{l-1,l}$, the transfer function of this layer according to Eq. (\ref{eq:Rouard_transfer_function_single}) is used leading to the combined transfer function

\begin{equation}
    H_{\text{total}}^{2}(\omega) = r_{l-2, l-1} + \frac{t_{l-2, l-1}\,  t_{l-1, l-2}\, H_{l}(\omega)\,\text{exp} \big (\frac{2\text{i}\tilde{n_{l}}\omega d_{l}}{c} \big )}{1-r_{l-1, l-2}\, H_{l}(\omega)\, \text{exp} \big (\frac{2\text{i}\tilde{n_{l}}\omega d_{l}}{c} \big )}
    \label{eq:Rouard_transfer_function_double}
\end{equation}

for a two-layer sample and so on for further layers. In this way, Rouard's method enables to model an arbitrary number of layers depending on the material parameters \cite{krimi2016non}.

In this paper, the investigated SiC samples are 3-layer systems consisting of an epilayer, a buffer layer, and a substrate, as Fig. \ref{fig:Setup} shows. The material parameters for the static relative permittivity, the effective mass, and the charge carrier mobility, that are inserted into Eq. (\ref{eq:dielectric_function_Drude_N_dependend}), are based on literature values for the evaluation. However, the material parameters for 4H-SiC are not as precisely determined in literature as one might suppose. 

For the TDS evaluation, an analysis of the error propagation based on the possible values of the material parameter according to \cite{burin2024tcad} leads to a variation of the result for $N_{\text{Epi}}$ of about 7\% to 15\% for $\epsilon_{\text{s}}$ in the range of 9.63 to 10.35, 19\% to 24\% for $m_{\text{eff}}$ between the value 0.33 corresponding to m$_{\text{eff,}\parallel}$ and 0.404 corresponding to $m_{\text{eff,}\perp}$, 19\% to 27\% when using $d_{\text{Epi}}$ received from the FTIR measurements assuming $\epsilon_{\infty}$ in the range of 6.25 and 7.56, and 19\% to 28\%  for various parameters for the Caughey-Thomas formula for $\mu(N)$ found in \cite{burin2024tcad} and the works cited there. However, the exact deviation always depends on the doping density of the wafer under test due to the shifts on a logarithmic scale described by percentage values. Therefore, a rounded error range of 20\% is assumed for the TDS results. These discrepancies can come from different measurement techniques, calculations, or assumptions and show that research concerning these parameters is still ongoing \cite{burin2024tcad}.

According to Fig. \ref{fig:Setup}, both the angle of incidence and the angle of reflection of the terahertz radiation are almost parallel to the c-axis, leading to the electric field oriented perpendicularly. When taking refraction into account, the angle of propagation inside the material is even smaller than the angle of incidence. Therefore, a small-angle approximation is justified and the perpendicular values for these material parameters are used. The error of the approximation to use the values perpendicular to the c-axis is smaller than 1\% for each parameter individually as well as for the resulting effect on the determined charge carrier density value. With these considerations in mind, the following values for the material parameters are used in this work:

\begin{equation}
    \epsilon_{\text{s,}\perp} = 9.91 \;\;\;,\;\;\;  m_{\text{eff,}\perp} = 0.404 \;\;\;,\;\;\; \mu_{\perp} = \frac{910 - 40}{ 1 + \big (\frac{N}{2\times10^{17}} \big )^{0.76}} + 40 \; \frac{\mathrm{cm^2}}{\text{Vs}} .
    \label{eq:SiC_material_parameters}
\end{equation}

While a well-arranged and in-detail overview of the various literature values for the parameters can be found in \cite{burin2024tcad}, we want to highlight the references of the original works, the parameters in Eq. (\ref{eq:SiC_material_parameters}) are taken from: For $\epsilon_{s}$, in \cite{naftaly2016silicon} a Sellmeier-equation-based analysis of the refractive index in the terahertz region makes this reference a suitable one for our study. The anisotropy of the drift mobility and the perpendicular effective mass is analyzed in \cite{ishikawa2023experimental}. However, the Hall measurements, conducted in this paper, were only measurements of few samples and are therefore not used as a reference for the mobility. Instead, another review-like study evaluated measurements of the mobility of numerous other works. Hence, these results are used for $\mu_{\perp}$ \cite{stefanakis2014tcad}. The mobility also depends heavily on the layer thickness \cite{masuda2000determination}. Therefore, with the main goal to correctly model a thin epilayer, a parameter for a mobility with lower values than in many other works is assumed \cite{burin2024tcad}. The temperature dependence is neglected because all measurements are recorded at room temperature.

Furthermore, the layer thickness is assumed to be constant across each sample. The combined thickness of the epilayer and the buffer layer is determined from FTIR measurements, which is the established technique for this purpose. 
% The evaluation of the spectra is done by applying a fast Fourier transform to the spectra roughly the range of wavenumbers from 3600 cm$^{-1}$ to 2500 cm$^{-1}$. The thickness is calculated by dividing the results by the refractive index and by 2 due to two passages through the layers. 
To retrieve the thickness from the FTIR measurement, the relative permittivity is calculated using the Sellmeier equations from \cite{wang20134h} and used to determine the thicknesses of the samples.

With the material parameters from Eq. (\ref{eq:SiC_material_parameters}), the sample thickness and Rouard's method, the interaction of the terahertz radiation with the material is modeled. In the following evaluation, a simulation of the expected material response and its effect on a reference measurement is calculated and compared to the measurement. This is done in the frequency domain to include dispersion. In a two-stage optimization process, the correlation between the measurement and the simulation is optimized and the sought-after charge carrier density is determined.

\section{Simulation}

To characterize the range of charge carrier densities, the used TDS setup is sensitive to, simulations based on the described model, material, and system parameters are computed for various combinations of epilayer thicknesses and charge carrier densities. 

\begin{figure}[htbp!]
\centering
\centering\includegraphics[width=0.75\textwidth]{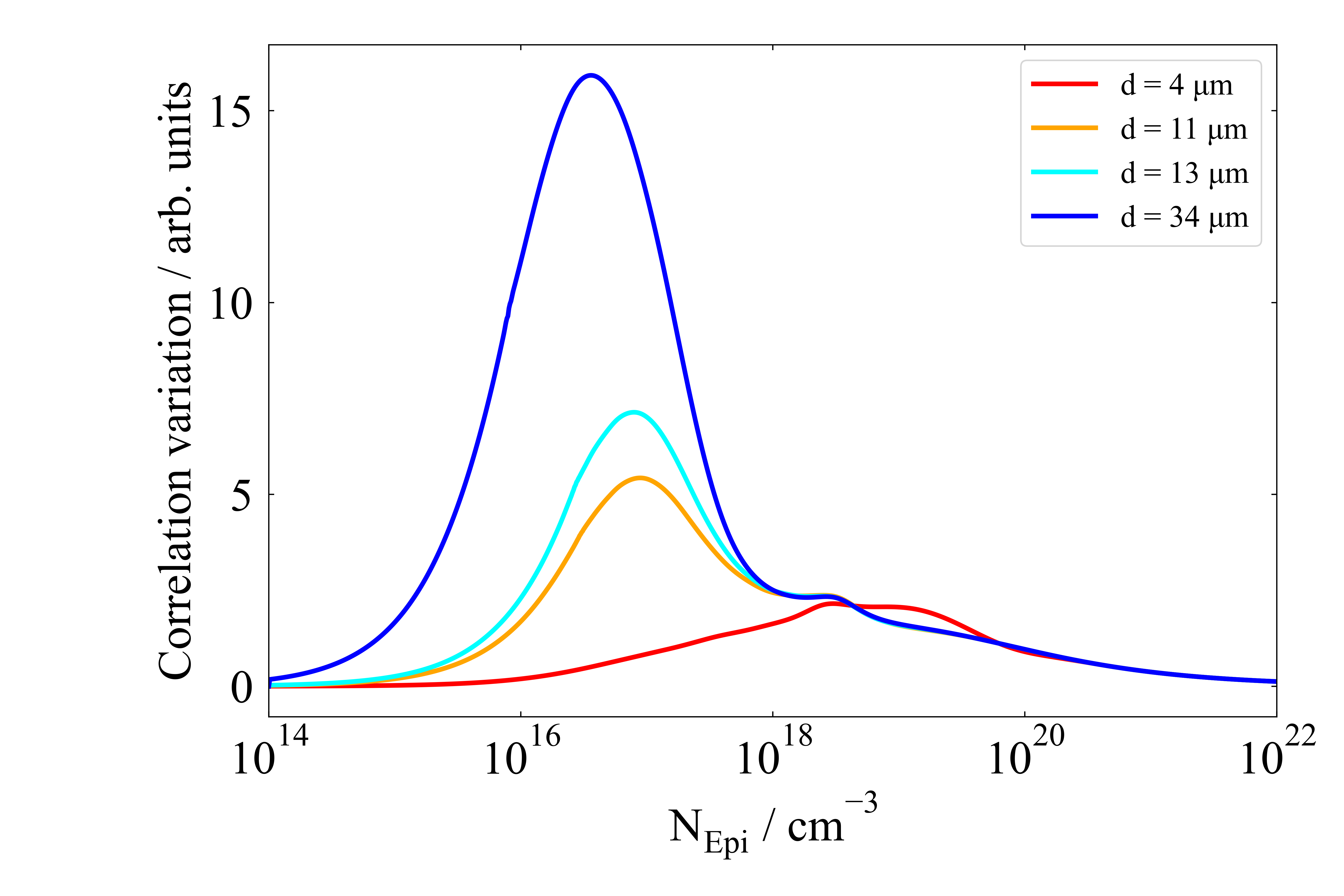}
\caption{Variations of the correlation of simulated pulses after simulated interaction with the modeled 4H-SiC samples over a wide range of charge carrier densities and for different epilayer thicknesses. The correlation variation refers to the quantitative change corresponding to the next $N_{\text{Epi}}$ value simulated and can therefore be understood as the sensitivity of the measurement result to variations of $N_{\text{Epi}}$.} 
\label{fig:Simulation}
\end{figure}

Fig. \ref{fig:Simulation} shows the computed correlation variation for four different modeled 3-layer stacks of 4H-SiC according to the setup in Fig. \ref{fig:Setup}. The first step to obtain the graphs is to simulate the interaction of an idealized terahertz pulse with a sample as described in chapter 4 for charge carrier density values in the range from 10$^{14}$ cm$^{-3}$ to 10$^{22}$ cm$^{-3}$. By comparing each of these simulations with the analogous simulation for the neighboring value of $N_{\text{Epi}}$, the correlation variation results from their root-mean-square deviation. It can therefore be understood as a measure of the derivative of the correlation. By varying the charge carrier density over a wide range and calculating these simulations for multiple realistic epilayer thicknesses, depicted with different colors, a well-founded assessment of whether a certain epilayer of a sample can be characterized is possible. 

Therefore, the amplitudes of the curves in Fig. \ref{fig:Simulation} correspond to the expectable dynamic in the terahertz signal in a certain doping range. While for thicker samples the curves not only become steeper and reach higher amplitudes, they also start to rise for smaller doping levels. Due to the longer interaction length with free charge carriers, the terahertz radiation is affected more by even a small extinction rate in the lightly doped material. Furthermore, for thicker layers, multireflections can also be resolved in the time domain leading to additional echo pulses, which yields more distinction criteria for the waveforms. For the extreme case of hundreds of microns thick wafers, the capability of TDS has been characterized for silicon in our previous study \cite{hennig2024wide}.

For the epilayers becoming thinner, a shift of the peak towards higher doping levels can be observed because of the shorter length for interaction with free charge carriers provided. The higher the charge carrier density is, the higher the extinction rate becomes until the sample appears mostly opaque depending on the frequency. 
%For which $N_{\text{Epi}}$ the transition between transparency and opacity occurs, also depends on the thickness of the epilayer due to the light propagating twice through the layer in the reflection setup. 
When the epilayer is mostly opaque, the Fresnel reflections from the air-SiC interface, depicted in Fig. \ref{fig:SiC_Drude_model} (b), play the dominant role in the interaction between pulse and sample. In this extreme case, only the long-wavelength components of the pulse have the potential to be partially transmitted and probe the whole epilayer to the buffer layer. For highly doped samples, this is only possible for thin layers. In this case, however, the peaks of the reflection from the front surface and the interface between epilayer and buffer layer cannot be resolved separately in the time domain anymore. Also, the respective Fabry-Pérot dips in the recorded spectrum become further apart or disappear from the relevant frequency range, respectively.

The range of characterizable charge carrier densities with TDS can be compared to that of terahertz time-domain ellipsometry, a similar measurement technique based on altering the polarization of the incident terahertz radiation under certain angles. To our knowledge, no SiC epilayers have yet been characterized with that technique, but highly doped 4H-SiC substrates have been \cite{agulto2025wafer}. While there are no publications of full simulations of the interaction between the terahertz radiation and the material, the expected sensitivity range is estimated to be roughly 10$^{16}$~cm$^{-3}$ to 10$^{21}$~cm$^{-3}$ \cite{agulto2025wafer, iwamoto2023characterization} . This is in good agreement with our simulations in Fig. \ref{fig:Simulation}, which is to be expected given that the same wavelength is used in both techniques.

\section{Measurement results}

\begin{figure}[htbp!]
\centering
\includegraphics[width=0.49\textwidth]{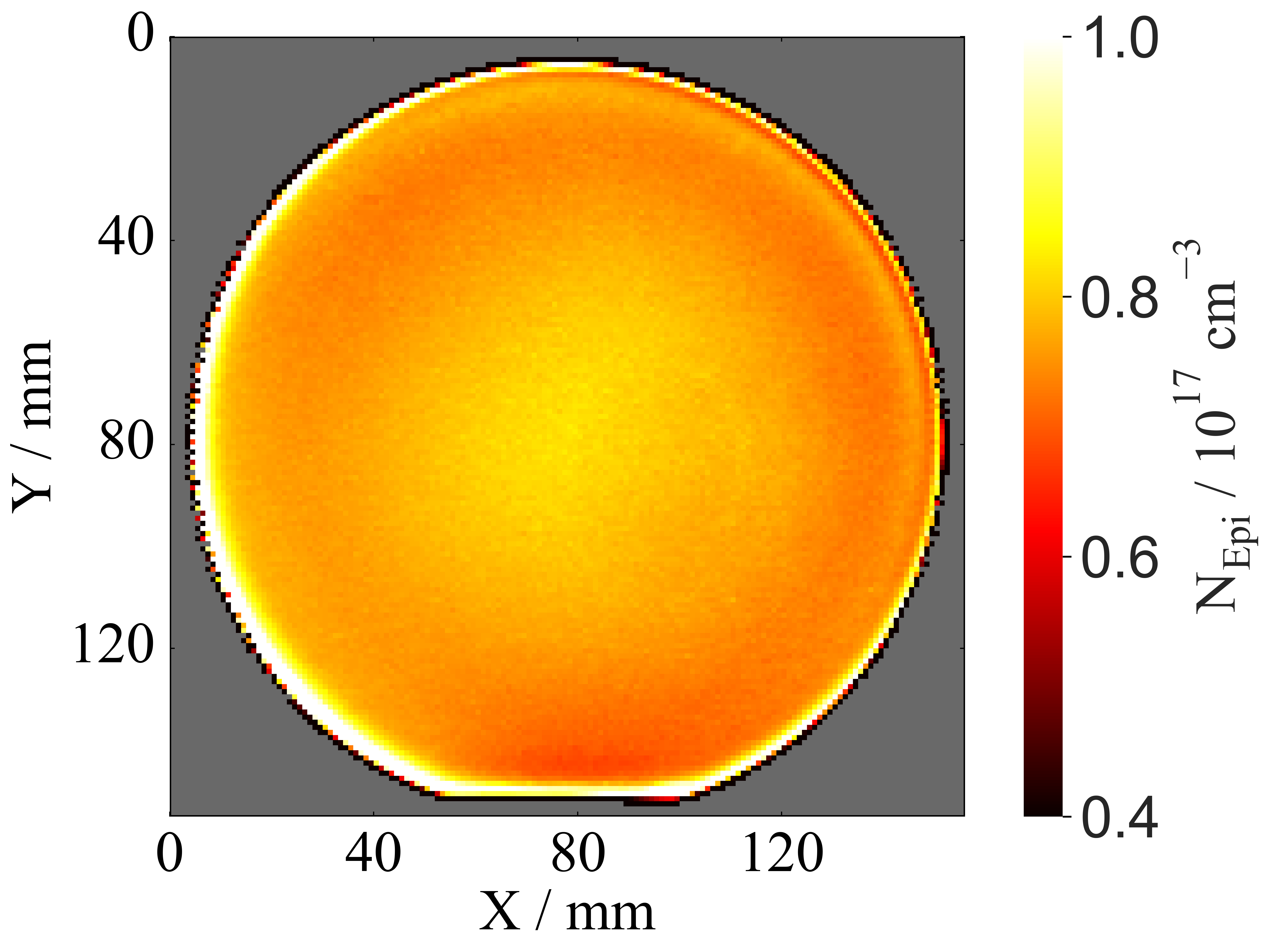}
\put(-190,120){(a)}
\put(-160,127){\textcolor{orange}{\large $\boldsymbol{\times}$}}
\put(-76,125){\small\textcolor{white}{11.1 µm}}
\hspace{0.1cm}
\includegraphics[width=0.49\textwidth]{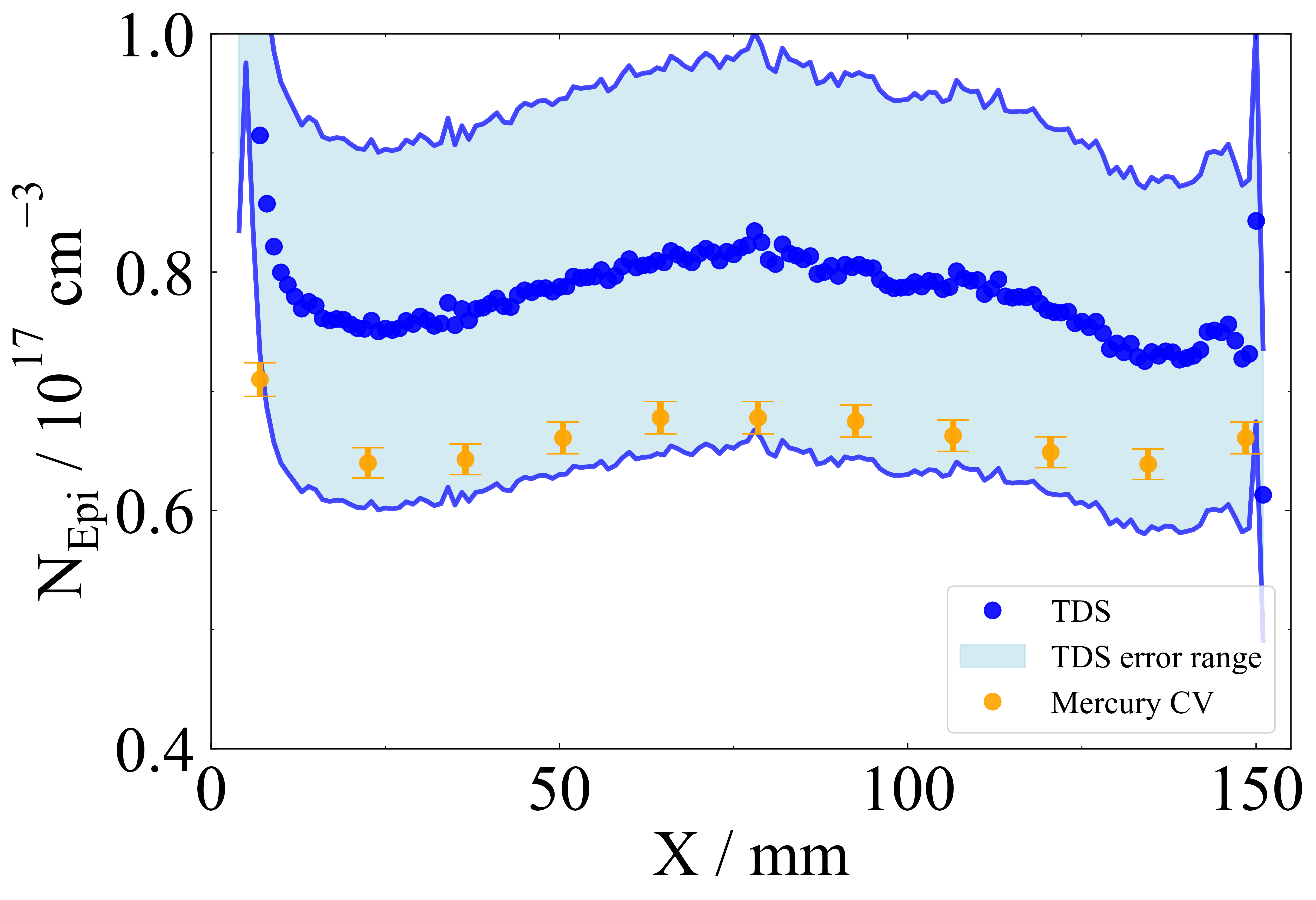}
\put(-185,120){(b)}
\vspace{0.01cm}

\includegraphics[width=0.49\textwidth]{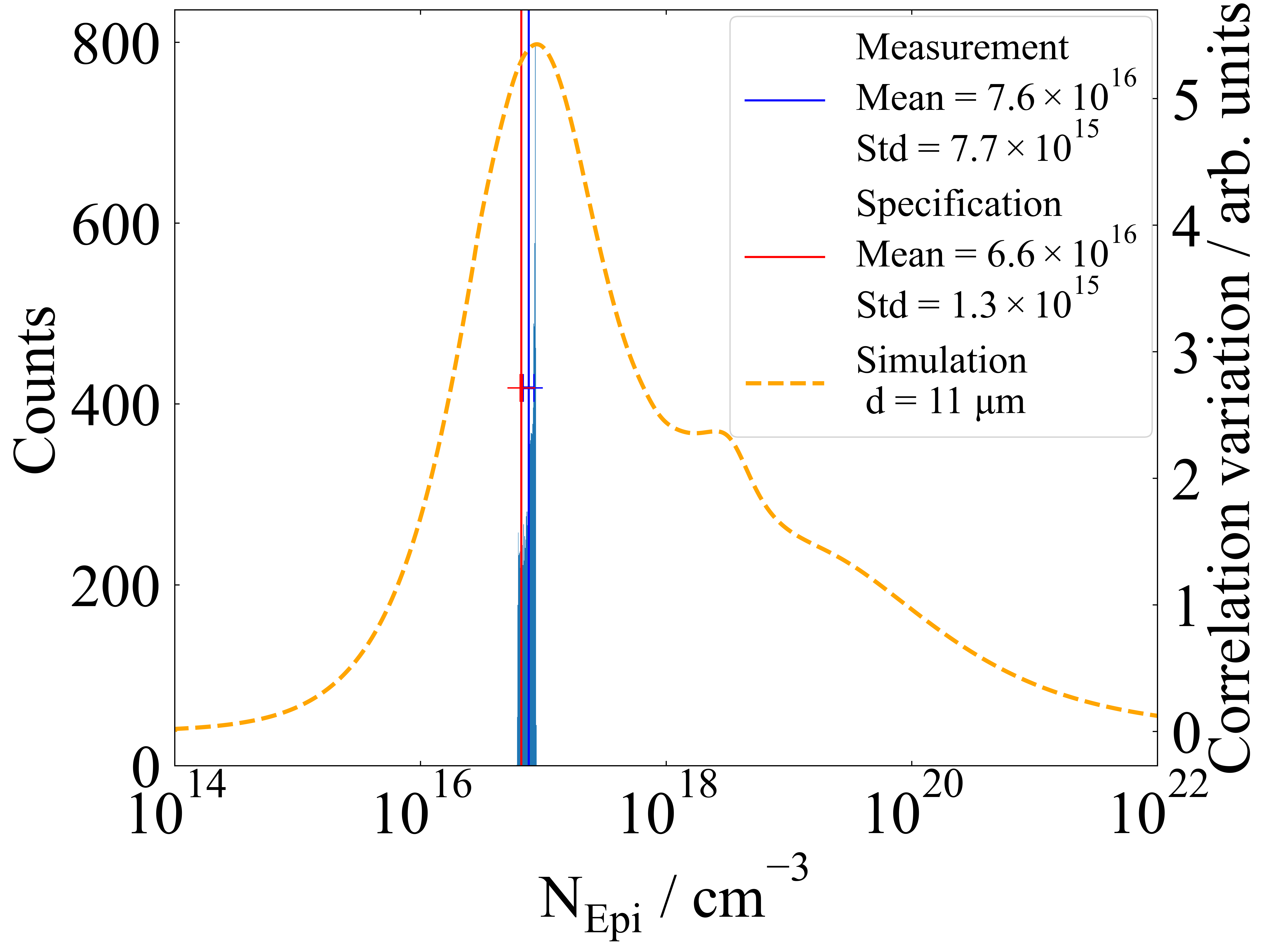}
\put(-190,100){(c)}
\hspace{0.1cm}
\includegraphics[width=0.49\textwidth]{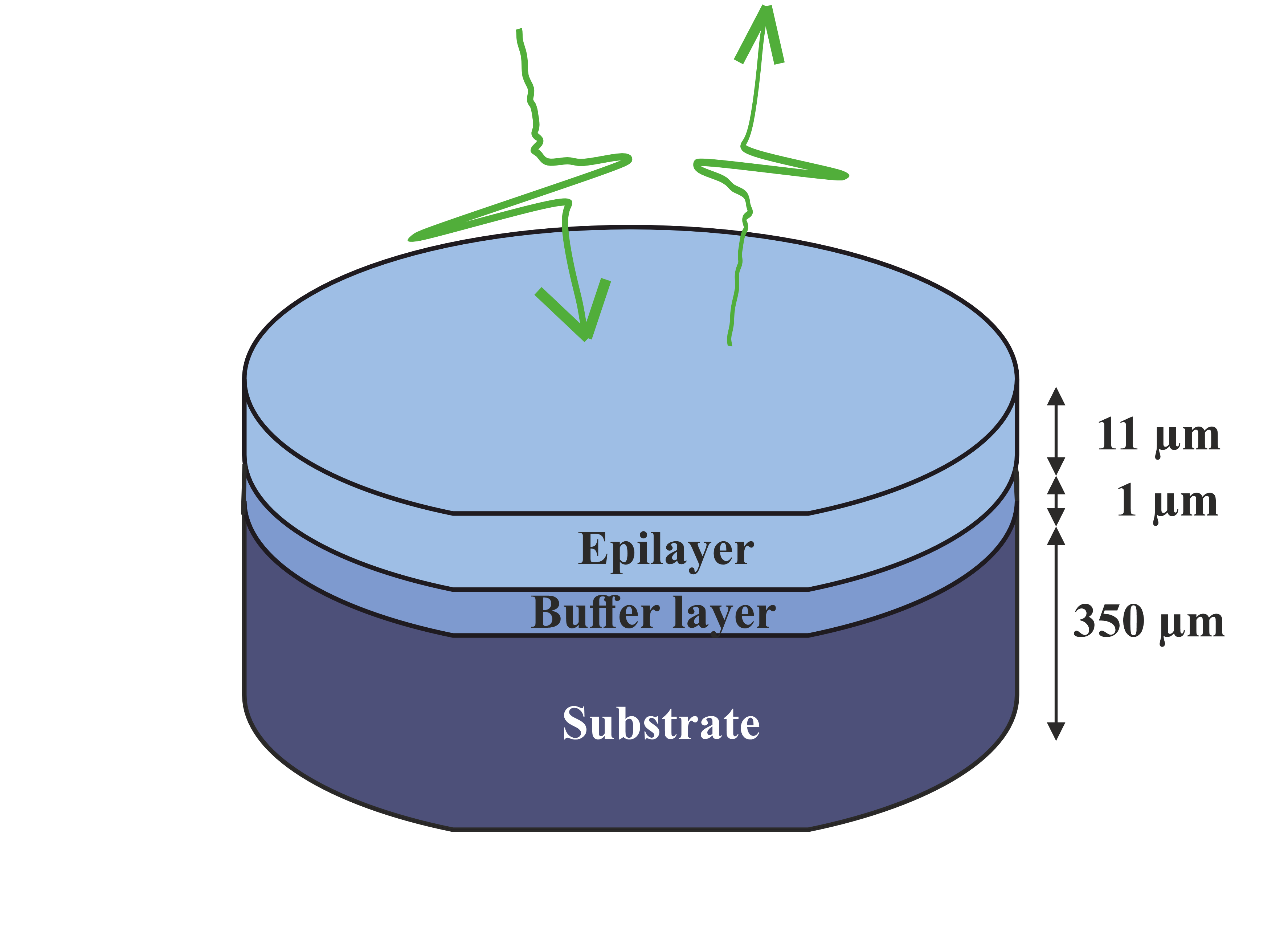}
\put(-185,100){(d)}
\caption{(a) $N_{\text{Epi}}$ determined from an X-Y scan of TDS measurements for a 3-layer 4H-SiC wafer, (b) a line scan of the same wafer across the X axis, (c) the histogram corresponding to the results of the X-Y scan compared to the mCV specification and the simulation corresponding to the epilayer thickness of this sample, and (d) a schematic of the sample and the incident and reflected terahertz pulses.}
\label{fig:07SY}
\end{figure}

Multiple SiC samples were measured, corresponding to the four epilayer thicknesses used in Fig. \ref{fig:Simulation}. In Fig. \ref{fig:07SY}, the X-Y scan of one of them is evaluated in detail. Fig. \ref{fig:07SY} (a) shows the distribution of the charge carrier density of the epilayer across the whole wafer based on the evaluation described in chapter 4. An additional correction with a plane was applied to account for a systematic gradient across the sample. This could possibly be due to a slight vertical misalignment of the sample in the measurement setup and can be observed for all samples, being more prominent for thinner epilayers. A small, circularly distributed variation can be observed, as expected for a homogeneously grown epilayer. Towards the center of the wafer, $N_{\text{Epi}}$ is higher and decreases towards the edge from about 8.4$\times$10$^{16}$~cm$^{-3}$ to about 7.5$\times$10$^{16}$~cm$^{-3}$. 

This behavior can also be observed in the cross section in Fig. \ref{fig:07SY} (b) across the diameter of the wafer in Y direction along the whole X axis. The results of the TDS measurement are represented by blue marks and the mCV measurements, which serve as a reference in this paper, by orange marks. The same qualitative and quantitative trend across the line scan clearly shows the agreement between both methods. The offset of approx. 16\% can be explained by possible deviations in the assumptions of the material parameters in each measurement, as explained in chapter 4. In addition, calibration errors or the uncertainty of other relevant measurement parameters, such as the contacting area of the mercury, can play a role as possible systematic errors of the reference method. However, as a minimum uncertainty the sigma/mean deviation of the mCV measurements with approx. 2\% is indicated by error bars for these data points.

% Therefore, a rounded error of 20\% is assumed for the TDS results. With a calibration, this could be reduced further. The error in reproducibility of this measurement is below 2\% for more than 5 averages of separate measurements at the same position, and below 1\% for more than 20 such measurements averaged. For the 100 averages-based results shown here, a reproducibility of 0.7\% is achieved. 

The advantage in measurement speed is obvious when comparing the needed approx. 20 minutes for 25 data points of the mCV measurement and approx. 70 minutes for 17609 measurements -- excluding the discarded data points next to the wafer, which were recorded due to rectangular shape of the X-Y scan type, shown as the grey area in Fig 5 (a) -- resulting in a measurement rate being more than 200 times higher.

Fig. \ref{fig:07SY} (c) shows the histogram of the raster scan of Fig. \ref{fig:07SY} (a) as blue bars, neglecting the edge values. With a mean value $N_{\text{Epi}}$ and standard deviation of $(7.6\pm0.8)\times$10$^{16}$~cm$^{-3}$, a small sigma/mean deviation of approx. 10\% across the whole wafer is measured. The measurement and the red-marked specification also show an overlap in their standard deviation. In the background, the dashed line shows the theoretical graph of the correlation variation for the thickness of this samples according to Fig. \ref{fig:Simulation}. This shows that this sample lies in an area of high sensitivity of this method for the given epilayer thickness. Fig. \ref{fig:07SY} (d) schematically shows the sequence and thicknesses of the layers of this sample and the reflection geometry of the pulsed measurement.

\begin{figure}[htbp!]
\centering
\includegraphics[width=0.49\textwidth]{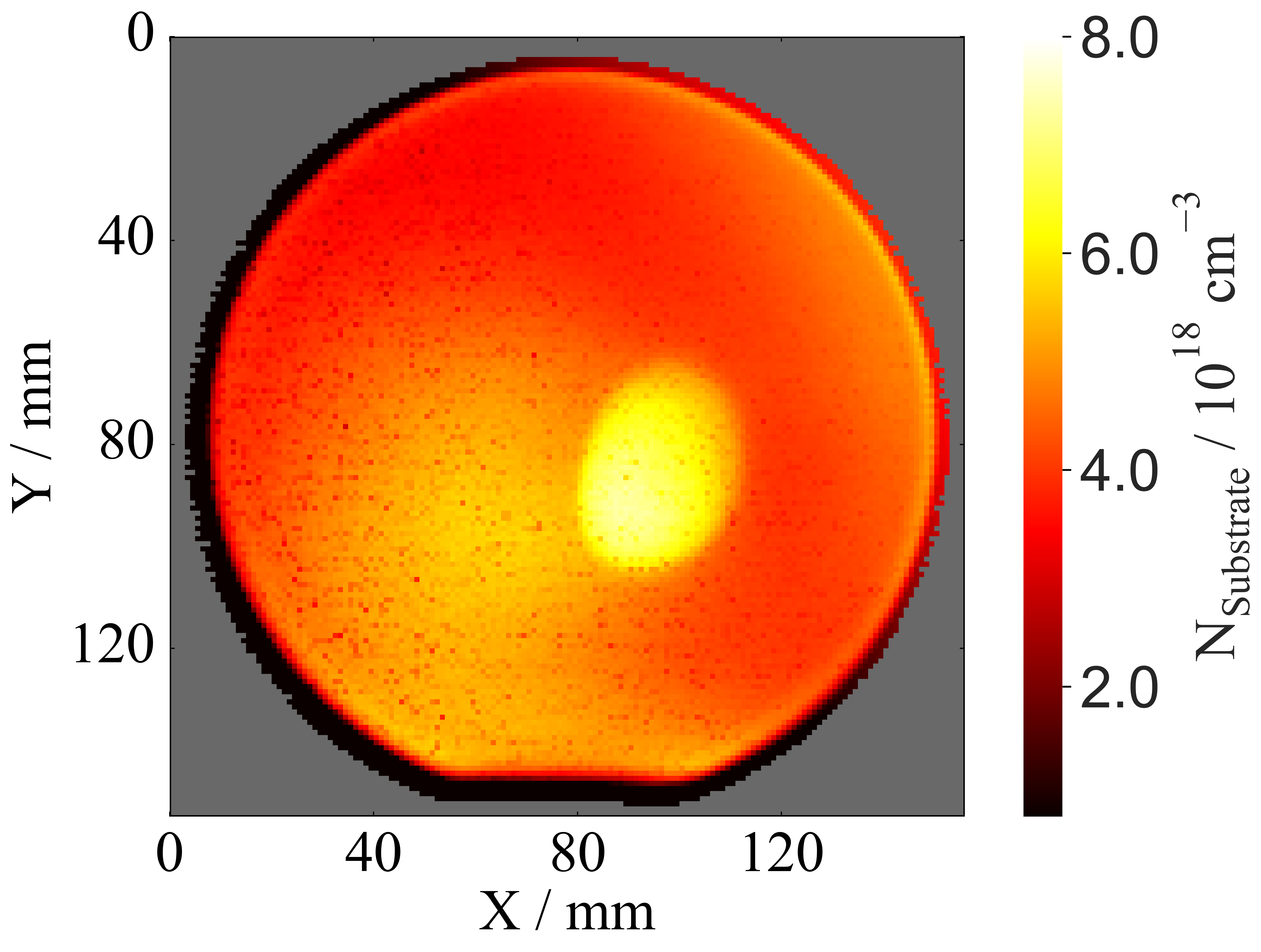}
\put(-180,120){(a)}
\put(-160,127){\textcolor{black}{\large $\boldsymbol{\times}$}}
\put(-80,128){\small\textcolor{white}{Substrate}}
\put(-80,22){\small\textcolor{white}{front side}}
\hspace{0.1cm}
\includegraphics[width=0.49\textwidth]{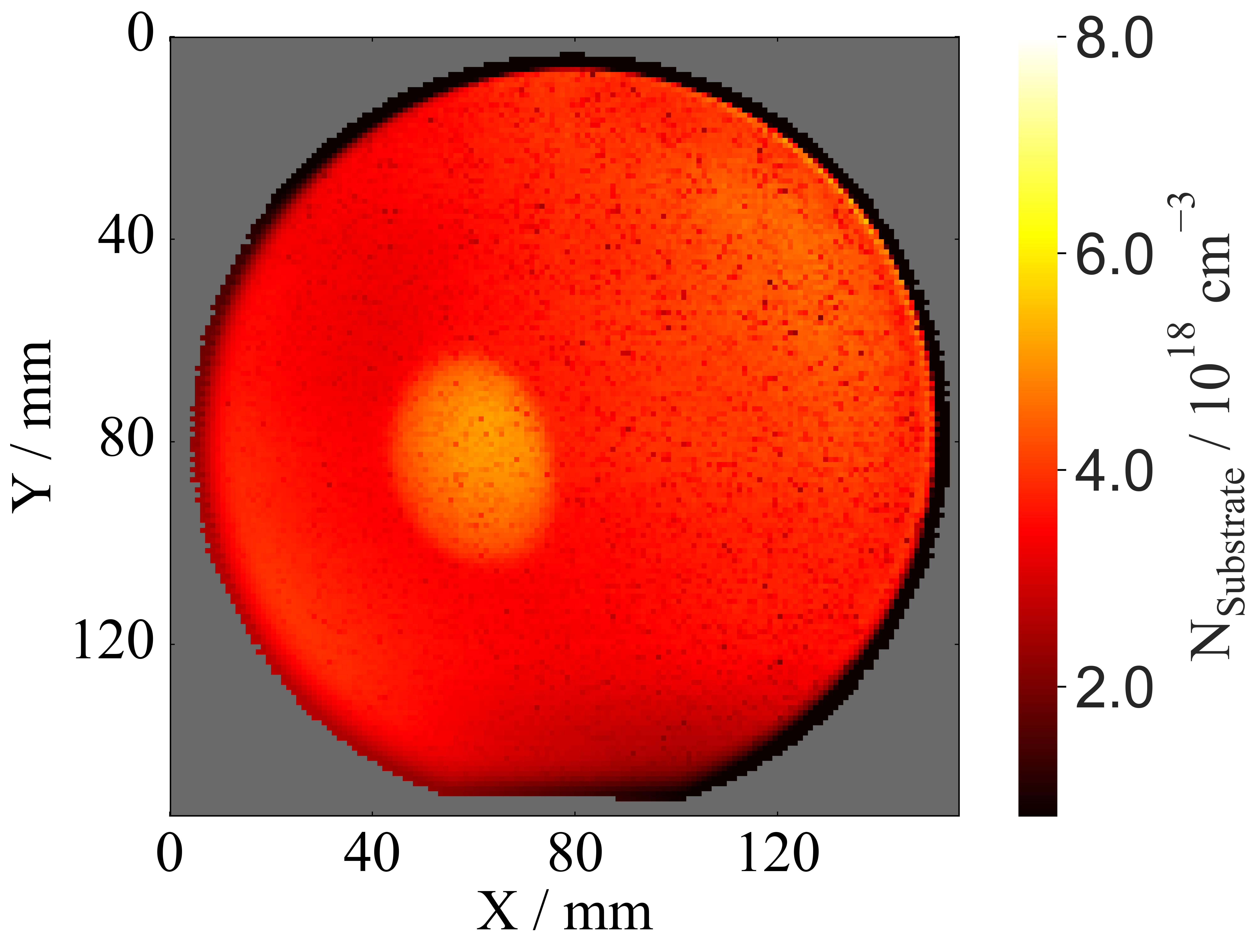}
\put(-180,120){(b)}
\put(-160,127){\textcolor{black}{\large $\boldsymbol{\times}$}}
\put(-80,128){\small\textcolor{white}{Substrate}}
\put(-79,22){\small\textcolor{white}{back side}}
\caption{X-Y scan results of the charge carrier density of the substrate for the same 3-layer 4H-SiC wafer as in Fig. \ref{fig:07SY}, (a) from the same epilayer-side measurement and (b) measured independently directly from the substrate side.}
\label{fig:Substrate}
\end{figure}

Another advantage of the TDS method compared to surface-based methods such as mCV is the fact that the stack of layers is probed completely. Therefore, from the same X-Y scan resulting in the charge carrier densities of the epilayer in Fig. \ref{fig:07SY} (a), the charge carrier densities of the substrate are also retrieved, as shown in Fig. \ref{fig:Substrate} (a). Again, a mostly homogeneous structure can be observed, except for the distinct facet area of the 4H-SiC substrate. This facet area is typical for doped 4H-SiC substrates. It originates from the PVD growth process of the material, where in the <0001> direction more nitrogen is incorporated near the center for undoped samples \cite{li2005nonuniformities}. Also, for intended N-doped samples, a facet appears due to increased incorporation of impurities during the growth process at the {0001} facet, which is growing fast and in a spiral manner. Due to the generally slower growth rate along the <0001> direction, the dopant concentration is typically increased by 20\%--50\% in the facet region compared to the outer region of the wafer \cite{kimoto2014fundamentals}. The exact position of the facet as well as its extent across the wafer differ between different suppliers of the substrates. In Fig. \ref{fig:Substrate} (a), a facet can be observed in the substrate, measured from the epilayer side. Fig. \ref{fig:Substrate} (b) shows a direct measurement of the substrate side after the wafer is turned around. The facet position is mirrored due to the sample being flipped. Apart from that, the same structure of the sample can be observed including the facet. For both plots, a small gradient across the whole measurements can still be observed, even after correction. This is most likely due to a small tilt of the sample in relation to the measurement head. The mean and standard deviation for both evaluations are (a) $(4.6 \pm 0.8 )\times10^{18}$~cm$^{-3}$ and (b) $(3.7 \pm 0.4)\times$10$^{18}$~cm$^{-3}$, deviating less than 20\%.
%(a) 4.6$\times$10$^{18}$ $\pm$ 8.0$\times$10$^{17}$ cm$^{-3}$ and (b) 3.7$\times$10$^{18}$ $\pm$ 4.3$\times$10$^{17}$ cm$^{-3}$, deviating less than 20\%.

% \definecolor{meineFarbe}{rgb}{0.1, 0.6, 0.2}

\begin{figure}[htbp!]
\centering
\includegraphics[width=0.49\textwidth]{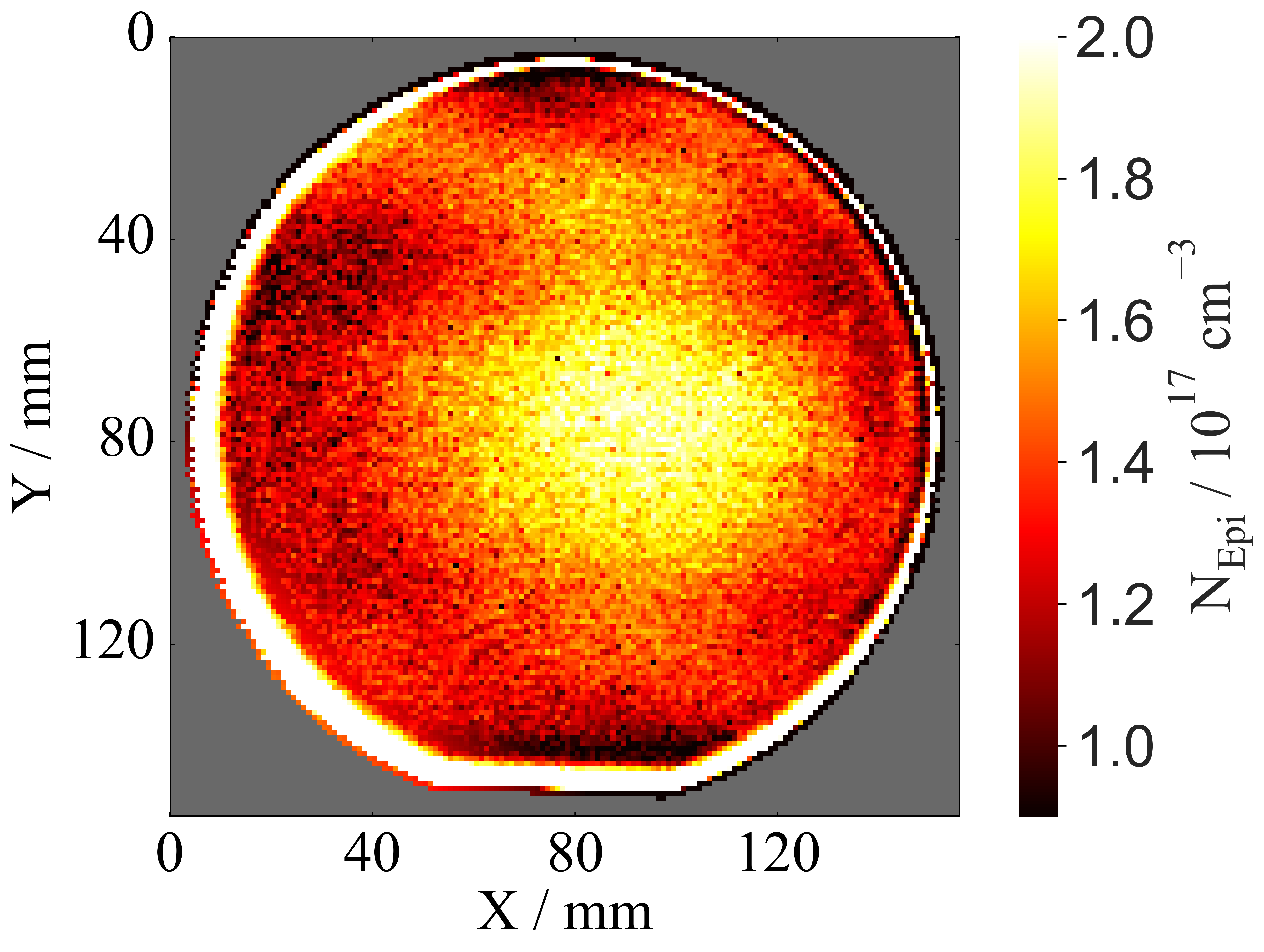}
\put(-180,120){(a)}
\put(-160,127){\textcolor{red}{\large $\boldsymbol{\times}$}}
\put(-72,125){\small\textcolor{white}{4.1 µm}}
\hspace{0.1cm}
\includegraphics[width=0.49\textwidth]{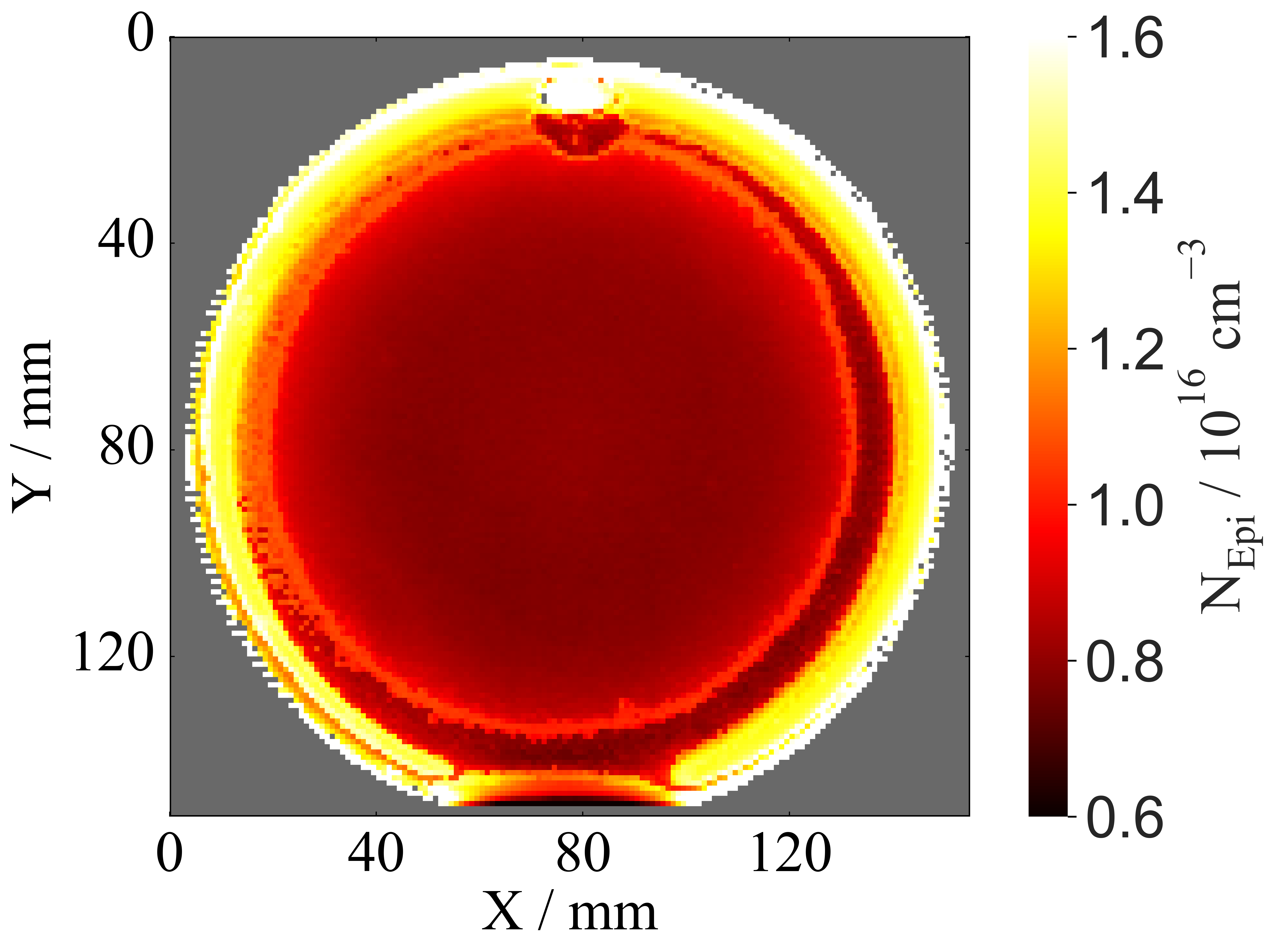}
\put(-180,120){(b)}
\put(-160,127){\textcolor{blue}{\large $\boldsymbol{\times}$}}
\put(-76,125){\small\textcolor{white}{32.9 µm}}
\vspace{0.01cm}

\includegraphics[width=0.49\textwidth]{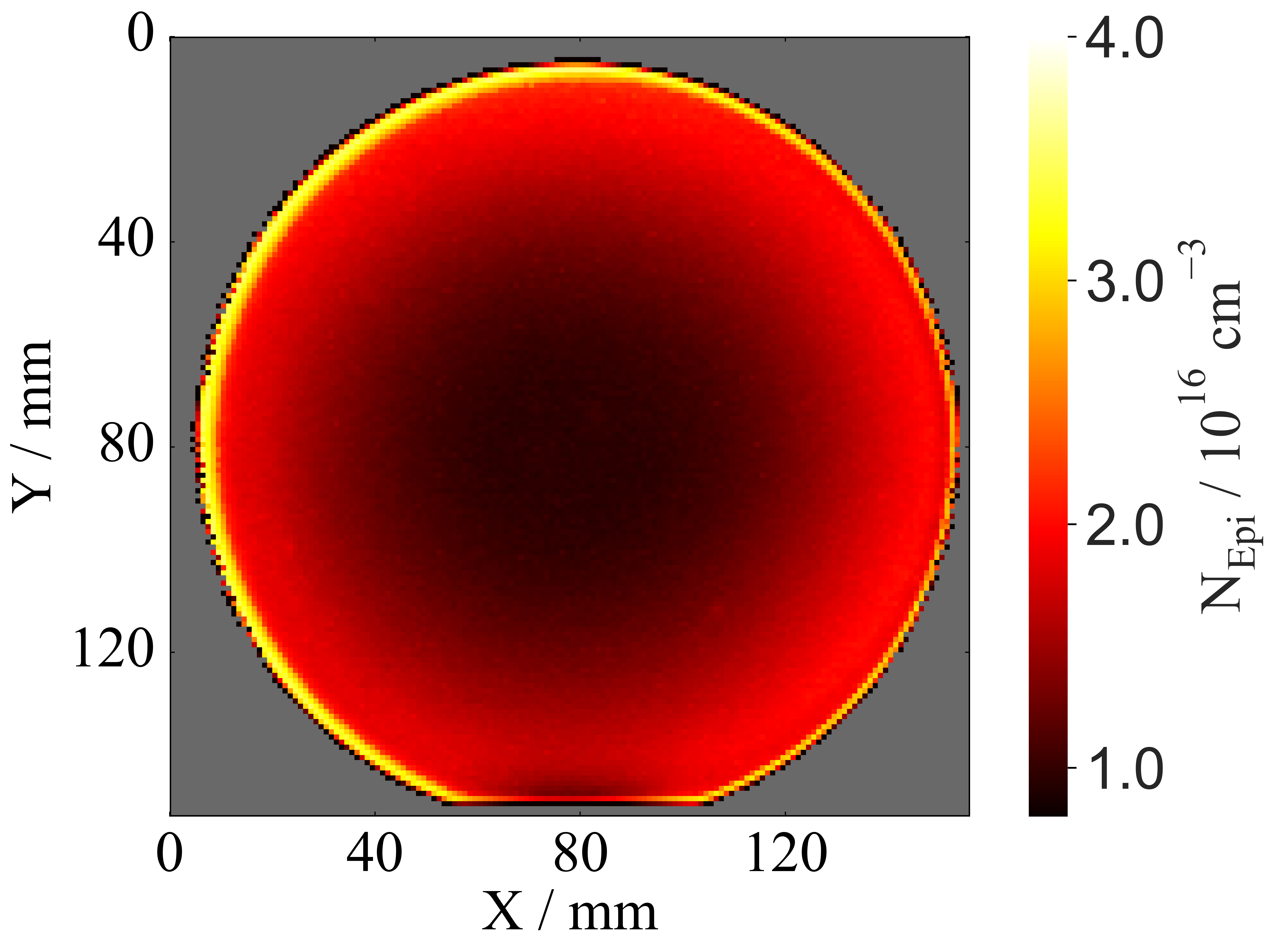}
\put(-180,120){(c)}
\put(-160,127){\textcolor{cyan}{\large $\boldsymbol{\times}$}}
\put(-76,125){\small\textcolor{white}{13.7 µm}}
\hspace{0.1cm}
\includegraphics[width=0.49\textwidth]{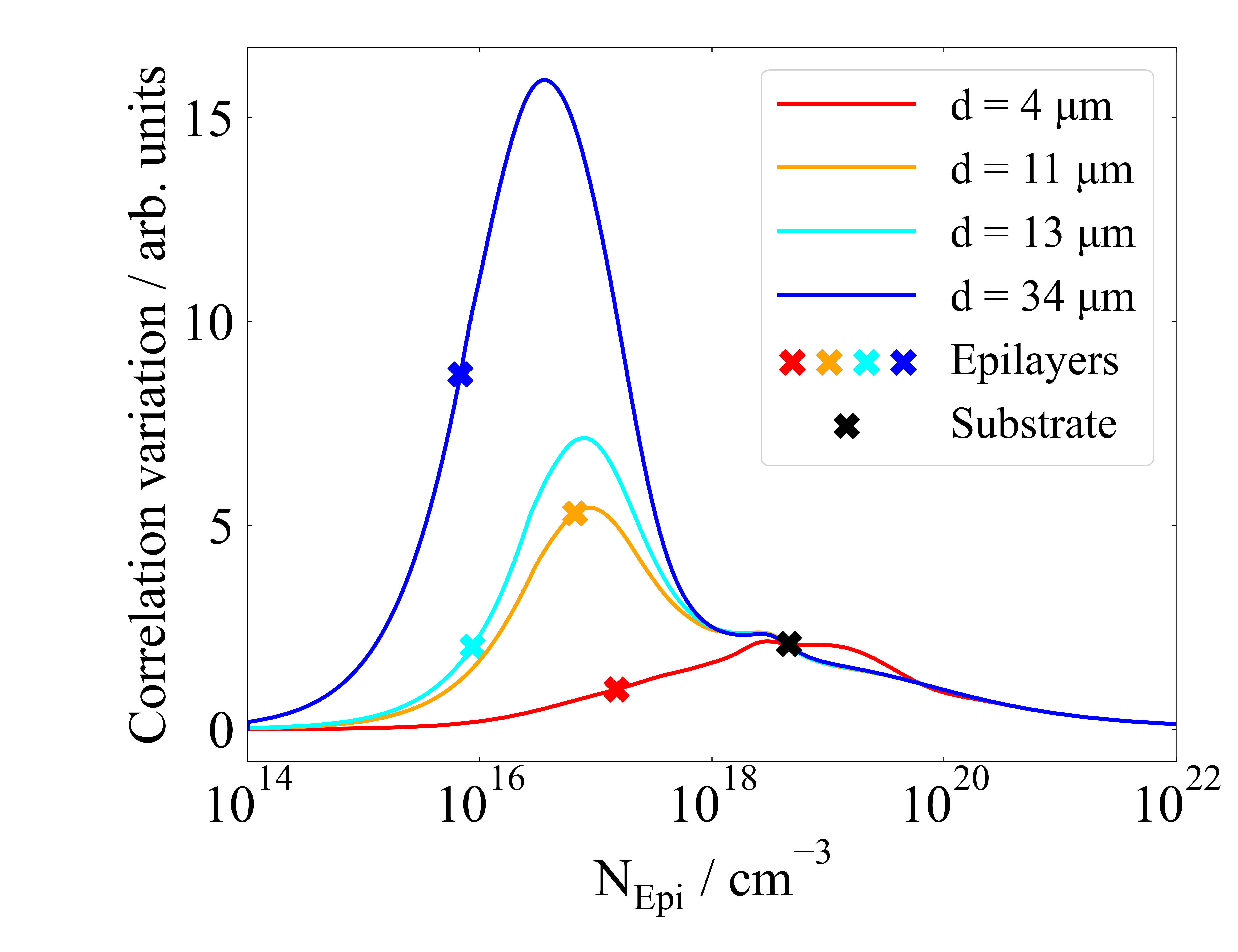}
\put(-145,120){(d)}

\caption{(a) - (c) Determined $N_{\text{Epi}}$ values for X-Y scans of three further 3-layer 4H-SiC samples and (d) simulations of correlation variation according to Fig. \ref{fig:Simulation} with marked $N_{\text{Epi}}$ and $d_{\text{Epi}}$ positions of the analyzed samples color-coded to their respective corresponding simulation  as well as the mark inside the maps. The epilayer thickness is given inside the maps as well. The black mark refers to the substrate, while the colored ones represent different epilayers.} 
\label{fig:Further_SiC_samples}
\end{figure}
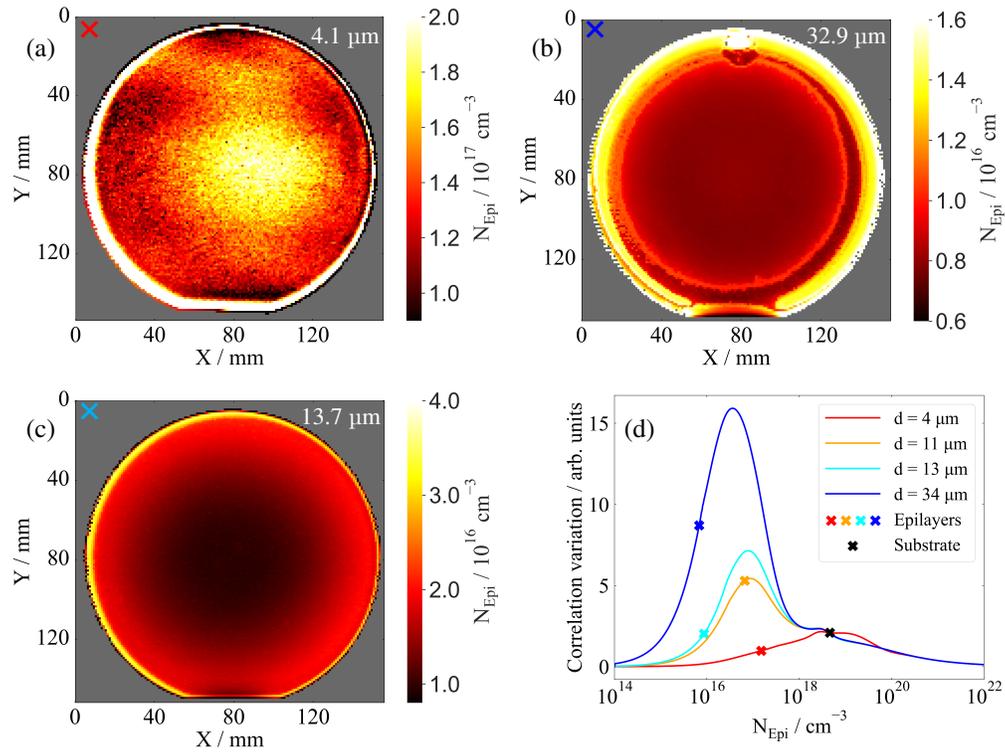

Further samples structured analogously to the one examined in Fig. \ref{fig:07SY} are measured and evaluated accordingly. The results of the scans of the respective $N_{\text{Epi}}$ values are shown in Fig. \ref{fig:Further_SiC_samples} (a) - (c). All samples exhibit a centrosymmetric gradient, either with increasing or decreasing doping level towards the center, except for the one shown in Fig. \ref{fig:Further_SiC_samples} (b). The latter shows a more homogeneous distribution. Especially for Fig. \ref{fig:Further_SiC_samples} (c), the gradient can be explained at least partially with a gradient in thickness that is neglected by assuming $d_{\text{Epi}}$ to be constant. Therefore, the gradient in $N_{\text{Epi}}$ is observed exaggeratedly. 

For the sample in Fig. \ref{fig:Further_SiC_samples} (a), it should be noted that because the epilayer is very thin with about 4 \textmu m, the thickness value of the entire drift layer measured with FTIR was assumed for the epilayer instead of subtracting the buffer layer thickness from it. For all samples, the observation is qualitatively in agreement with the mCV measurements. Fig. \ref{fig:Further_SiC_samples} (d) depicts the doping- and thickness-dependent simulations of the expected correlation variation according to Fig. \ref{fig:Simulation} with the values of $N_{\text{Epi}}$ and $d_{\text{Epi}}$ marked at the corresponding positions and in the corresponding colors of the curves. 

The samples' mCV specification and TDS results are given and compared in Tab. \ref{tab:specs_and_results}. A good agreement of the mean values across the whole X-Y scan for the TDS measurement and for both line scans for the mCV measurement can be found for the first two samples with a deviation well below 20\% and therefore within the systematic error. For the other two samples, the higher deviations could be due to a lack of information concerning $d_{\text{Buffer}}$ for the sample in Fig. \ref{fig:Further_SiC_samples} (b) or due to a comparably large gradient in thickness for sample Fig. \ref{fig:Further_SiC_samples} (c). With the simulations shown in Fig. \ref{fig:Further_SiC_samples} (d) in mind, the results of the measurements can also be explained in the way that Fig. \ref{fig:Further_SiC_samples} (a) is the most noisy measurement due to having a very low correlation variation. The reason for that is the very thin epilayer of about 4~\textmu m which poses a challenge for measurements with terahertz radiation with typical wavelengths of multiple 100~\textmu m. Because of the short interaction length between the terahertz radiation and the layer, the relative error is comparably high. The other samples show less noise due to being in a range with higher correlation variation. In agreement with the simulations, the sample in Fig. \ref{fig:Further_SiC_samples} (b) is one of the least noisy samples.

\begin{table}[htbp]
\caption{Specifications of the mCV measurements and TDS results for the investigated SiC epilayers}
\label{tab:specs_and_results}
\centering
\begin{tabularx}{1\textwidth}{ | >{\centering\arraybackslash}p{7mm} || >{\centering\arraybackslash}p{13mm} || >{\centering\arraybackslash}X  || >{\centering\arraybackslash}X || >{\centering\arraybackslash}p{14mm} |}
\hline
 & & {\textbf{mCV} } & {\textbf{TDS} } & Deviation\\
\hline\hline
\renewcommand{\arrayrulewidth}{10pt}
Fig. & $d_{\text{Epi}}$ / \textmu m & $N_{\text{Epi}}$  / cm$^{-3}$ &  $N_{\text{Epi}}$  / cm$^{-3}$ &  $\frac{\text{TDS}}{\text{mCV}}$ / \%\\ 
\hline\hline
\ref{fig:07SY} & 11.1 & $(6.6\pm0.1)\times$10$^{16}$ & $(7.6\pm0.8)\times$10$^{16}$ & 15.2\\
\hline
\ref{fig:Further_SiC_samples} (a) & \phantom{0}4.1 & $(1.5\pm0.02)\times$10$^{17}$ & $(1.4\pm0.7)\times$10$^{17}$ & \phantom{0}6.7\\
\hline
\ref{fig:Further_SiC_samples} (b) & 32.9 & $(6.8\pm0.3)\times$10$^{15}$ & $(8.5\pm0.9)\times$10$^{15}$ & 25.0\\
\hline
\ref{fig:Further_SiC_samples} (c) & 13.7 & $(8.7\pm1.1)\times$10$^{15}$ & $(1.3\pm0.3)\times$10$^{16}$ & 49.4\\
\hline
\end{tabularx}
\end{table}

Besides the dependence on the correctness and applicability of the material parameters taken from literature, a weakness of the model is that it does not take into account that the mobility of each layer might behave slightly differently due to the different layer thicknesses and should be modeled individually. However, to keep the model usable and applicable to different layer stacks with different thicknesses, we only use the same model for the charge carrier density-dependent mobility of all layers. Nevertheless, the good qualitative and quantitative agreement between the distribution and absolute values of $N_{\text{Epi}}$, determined from the TDS measurements and from the reference mCV measurements, shows that TDS is capable for certain combinations of $N_{\text{Epi}}$ and $d_{\text{Epi}}$ to determine the charge carrier density of 4H-SiC epilayers and substrates in a single, comparably fast and non-contact measurement with the applied model.

\section{Conclusion and Outlook}

In this work, we evaluated the possibilities of terahertz time-domain spectroscopy in reflection geometry to characterize the charge carrier concentration of 4H-SiC epi-wafers. Drude model-based simulations help define the range of charge carrier concentrations that are measurable depending on the thickness of the layer of interest. These simulations agree well with measurements of multiple samples of thin epilayers of thicknesses in the range of 4 \textmu m to 34 \textmu m. Charge carrier concentrations between approx. 8$\times$10$^{15}$~cm$^{-3}$ to 4$\times$10$^{18}$~cm$^{-3}$ are determined. The results for the $N_{\text{Epi}}$ values agree well with the established mercury capacitance-voltage measurements showing agreements in both the gradient of the charge carrier concentrations across the wafer as well as in the quantitative values. Furthermore, both the epilayer's and the substrate's charge carrier concentrations can be measured in a single measurement. The additional advantage of TDS of allowing much higher measurement rates, especially when using an ECOPS system, is utilized to create images of the distribution of the charge carrier concentration of complete wafers by X-Y scanning with a resolution of 1 mm. This demonstrates that TDS is a useful tool to characterize 4H-SiC epilayers and substrates over a wide range of charge carrier concentrations. Consequently, TDS appears to be a promising and capable technique in production and quality control of 4H-SiC to help identify inhomogeneities of wafers and satisfy the demand for SiC-based power devices in the future.

\section{Back matter}

% Back matter sections should be listed in the order Funding/Acknowledgment/Disclosures/Data Availability Statement/Supplemental Document section. An example of back matter with each of these sections included is shown below. The section titles should not follow the numbering scheme of the body of the paper. 

\begin{backmatter}
\bmsection{Funding}
This project is supported by the Federal Ministry for Economic Affairs and Climate Action (BMWK) on the basis of a decision by the German Bundestag.

\bmsection{Acknowledgment}
The authors thank Birgit Kallinger and Nikolas Schabert of the Fraunhofer IISB for providing further information concerning the reference measurements of the samples and giving insights into SiC sample preparations.

The authors thank Christina Jörg and Julian Schulz of the RPTU University Kaiserslautern-Landau for assistance with the FTIR measurements of the epilayers.

\bmsection{Disclosures}
The authors declare no conflicts of interest.

\bmsection{Data Availability Statement}
Data underlying the results presented in this paper are not publicly available at this time but may be obtained from the authors upon reasonable request.

% \bmsection{Supplemental document}
% A supplemental document must be called out in the back matter so that a link can be included. For example, “See Supplement 1 for supporting content.” Note that the Supplemental Document must also have a callout in the body of the paper.

\end{backmatter}

%%%%%%%%%%%%%%%%%%%%%%% References %%%%%%%%%%%%%%%%%%%%%%%%%

%%%%%%%%%% If using BibTeX:
\bibliography{bibliography}

\end{document}